\documentclass[final,aps,prx,twocolumn,english,superscriptaddress,floatfix,longbibliography,amsmath]{revtex4-1}
\usepackage{latexsym}
\usepackage{graphicx}
\usepackage{dcolumn}
\usepackage{bm}
\usepackage{comment}
\usepackage{mathrsfs}
\usepackage[colorlinks,linkcolor=DarkBlue,urlcolor=DarkBlue,citecolor=DarkBlue]{hyperref}
\usepackage[usenames,svgnames]{xcolor}

\newcommand{\etal}{\textit{et al.}}
\newcommand{\angstrom}{\text{\normalfont\AA}}

\begin{document}	
		

\title{Intertwined Orders and Electronic Structure in Superconducting Vortex Halos }
	
\author{Yi-Hsuan Liu}
\affiliation{Department of Physics, National Tsing Hua University, Hsinchu 30013, Taiwan}
\affiliation{Department of Physics, University of Virginia, Charlottesville, VA 22904, USA}

\author{Wei-Lin Tu}
\affiliation{Division of Display and Semiconductor Physics, Korea University, Sejong 30019, Korea}

\author{Gia-Wei Chern}
\affiliation{Department of Physics, University of Virginia, Charlottesville, VA 22904, USA}

\author{Ting-Kuo Lee}
\affiliation{Department of Physics, National Tsing Hua University, Hsinchu 30013, Taiwan}
\affiliation{Department of Physics, National Sun Yat-Sen University, Kaohsiung 80424, Taiwan}
\affiliation{Institute of Physics, Academia Sinica, Nankang 11529, Taiwan}
	
\date{\today}

\begin{abstract}
We present a comprehensive study of vortex structures in $d$-wave superconductors from large-scale renormalized mean-field theory of the square-lattice $t$-$t'$-$J$ model, which has been shown to provide a quantitative modeling for  high-$T_c$ cuprate superconductors. With an efficient implementation of the kernel polynomial method for solving electronic structures, self-consistent calculations involving up to $10^5$ variational parameters are performed to investigate the vortex solutions on lattices of up to $10^4$ sites. By taking into account the strong correlation of the model, our calculations shed new lights on two puzzling results that have emerged from recent scanning tunneling microscopy (STM) experiments. The first concerns the issue of the zero-biased-conductance peak (ZBCP) at the vortex core for a uniform $d$-wave superconducting state. Despite its theoretical prediction, the ZBCP was not observed in most doping range of cuprates except in heavily over-doped samples at low magnetic field. The second issue is the nature of the checkerboard charge density waves (CDWs) with a period of about 8 unit cells in the vortex halo at optimal doping. Although it has been suggested that such bipartite structure arises from low-energy quasiparticle interference, another intriguing scenario posits that the checkerboard CDWs originate from an underlying bidirectional pair-density wave (PDW) ordering with the same period. We present a coherent interpretation of these experimental results based on  systematic studies of the doping and magnetic field effects on vortex solutions with and without a checkerboard structure. The mechanism of the emergent intertwined orders within the vortex halo is also discussed. 
\end{abstract}

\maketitle
		
\section{Introduction}

\label{sec:intro}

The high-temperature superconductivity (SC) of cuprates is marked by the many ordering tendencies that either compete or coexist with the superconducting phase itself~\cite{Keimer15,Fradkin15,Martin1990,Daou09, Varma20}.  For example, charge density wave (CDW), either unidirectional or bidirectional, have been observed in all families of hole-doped high-temperature superconductors (HTSCs)~\cite{Tranquada1995,Kivelson03,Comin16}. The extensively studied stripe order in cuprates correspond to a unidirectional CDW, often accompanied by a spin density wave (SDW). The CDW order, which is intimately related to the intriguing pseudogap phase of cuprates, is short-ranged at zero field~\cite{Blackburn13,Ghiringhelli12,Canosa14,Tabis17}. Both the strength and correlation length of the CDW is enhanced by the magnetic field~\cite{Wu11,Wu13,Zhou17}. On the other hand, the onset of SC causes the reduction of the CDW amplitude, suggesting that these two orders are strongly intertwined with each other~\cite{Ghiringhelli12,Chang12}. The many unusual features of charge-density modulations in the pseudogap phase, such as the doping dependence of the modulation period, suggest that the CDW is a subsidiary order which results from a more fundamental pair-density wave (PDW) ordering. 

A PDW is a superconducting state in which the pairing order parameter varies periodically in space~\cite{Agterberg20}. The idea of a modulated pairing density was first introduced by Fulde and Ferrell and by Larkin and Ovchinnikov (FFLO) in a BCS model to overcome the Pauli limiting effect of a magnetic field~\cite{Fulde1964,larkin1965}. Recently, PDW states without any explicit breaking of the time-reversal symmetry have been proposed to exist in cuprates, especially in connection with the pseudogap phases~\cite{Chen04,Berg09,PALee14}. In particular, the PDW order  and its partial melting could lead to a variety of vestigial states including the CDW, possibly coexisting with an SDW mentioned above, and an unusual charge-$4e$ superconductivity~\cite{Berg09b}, among others. The rich PDW phenomenology seems to provide a natural explanation for the complexity of cuprate HTSC. Moreover, the scenario of decoupled layers of orthogonal planar PDW orders could account for the observed huge anisotropy of resistivity well above the nominal SC transition temperatures in La$_{2-x}$Ba$_x$CuO$_4$ at the 1/8 hole doping~\cite{Li07,Tranquada08,Berg07}.

Direct imaging of PDW orders has also been reported in hole doped Bi$_2$Sr$_2$CaCu$_2$O$_{8+x}$ (Bi2212) using atomic-resolution superconducting STM tips~\cite{Du20}. The amplitude of these modulations is characterized by a eight-unit-cell periodicity or wavevectors $\mathbf Q_x = \left(1/8, 0 \right) (2\pi / a)$ and $\mathbf Q_y = \left(0, 1/8 \right) (2\pi / a)$, where $a$ is the lattice constant. Simultaneous measurement of local density of states found electronic modulations with wavevectors $\mathbf Q_{x,y}$ and $2 \mathbf Q_{x, y}$~\cite{Du20}, which are consistent with the picture of a PDW coexisting with a uniform $d$-wave superconductivity~\cite{Agterberg15}. The nature of this SC state at zero magnetic field and its relationship with the PDW order, however, remains to be resolved. Given the complex nature of the cuprate superconductors with several energetically nearly degenerate ordered states, extrinsic effects such as impurity and disorder likely play an important role in the stabilization of this intriguing state with mixed SC and PDW ordering.

As the pairing order parameter is suppressed near a vortex core, magnetic-field induced vortices offer a fruitful platform to further investigate the subtle interplay between PDW and superconducting condensate. Indeed, for quite some time, a number of STM experiments have reported a nonuniform charge density and spectra inside the halo region surrounding the vortex core~\cite{Hoffman02,Matsuba03,Matsuba07,Hanaguri09,Hamidian15,Machida16}. A checkerboard pattern with a $4a \times 4a$ unit cell were observed in almost all samples. This checkerboard charge density modulation can also be viewed as a bi-directional CDW with wavevectors $2\mathbf Q_x$ and $2 \mathbf Q_y$, which are twice that of a fundamental PDW discussed above~\cite{Du20,Edkins19}. 
What is also striking is the bipartite electronic structure at the vortex core~\cite{Machida16}. At low energy  quasiparticle interference (QPI) dominates in the conductance while at higher energy, the energy-independent  CDW patterns emerge. 

Recently this vortex halo has been further studied and proposed to be a bidirectional  PDW by Edkins {\em et al}.~\cite{Edkins19}. The discrete Fourier transform of the conductance map of the halo reveals a modulation of the charge density $N(\mathbf r)$ with peaks at wavevector $\mathbf Q_{x, y}$, in addition to the previously observed $2 \mathbf Q_{x, y}$. Phenomenologically, this again suggests the coexistence of uniform SC condensate $\Delta_{\rm SC}$ and the PDW order $\Delta_{\mathbf Q}$ with wavevectors $\mathbf Q = \mathbf Q_x$ and $\mathbf Q_y$. In this scenario, a CDW order could result from couplings $\Delta^{\,}_{\mathbf Q} \Delta^*_{\rm SC}$ as well as $\Delta^{\,}_{\mathbf Q} \Delta^*_{-\mathbf Q}$, hence giving rise to charge modulations with a wavevector $\mathbf Q$ and $2 \mathbf Q$, respectively~\cite{Agterberg20}.

The recent STM experiments also reinvigorate a long-standing puzzle of cuprates. Due to the nodes of $d$-wave pairing order, it is expected and confirmed by theoretical calculations~\cite{ Wang1995,Franz1998} that a zero-biased conductance peak (ZBCP)~\cite{notezbcp} should be present at an isolated vortex center. Intriguingly, however, this peak has so far not been detected in most experiments. Instead, a sub-gap structure was reported~\cite{Machida16}. Several scenarios, e.g. the presence of a concomitant antiferromagnetic or SDW order~\cite{Zhu01,Franz2002} or an induced hidden $d_{xy}$ pairing ~\cite{Franz1998} in the halo region, have been proposed to explain this discrepancy. A consensus has yet to be reached among researchers. 

The issue of the vortex-induced ZBCP is further complicated by a recent experiment showing that, in contrast to sub-gap structure under higher magnetic field at $B \sim 3$T, there is a ZBCP at the vortex core under extremely low magnetic field ($B \sim0.16$ T) in an over-doped Bi2212 sample~\cite{Renner21}. In addition, at a hole doping of around $\delta=0.2$, the conductance map and spectra in the halo region of a $3$T vortex are shown to be very similar to that due to a checkerboard charge modulation in under-doped samples reported by others. However, the wavevectors associated with the checkerboard seem to disperse within an energy range of 3 to 10 meV.  This result seems to be related with the aforementioned bipartite structure found by Machida \etal~\cite{Machida16} for an optimally doped sample under 11.25~T magnetic field, yet with very different dopant density and field strength.

All these recent results posed several new interesting questions. Is the ZBCP only present in over-doped samples? Or is it suppressed by the presence of PDW or checkerboard state? Is the bipartite structure inside a vortex an intrinsic property of the PDW state? What role does the QPI play in the formation of checkerboard halo states? What is the mechanism for the inducement of the checkerboard pattern inside a vortex?
To answer these questions, one must examine the vortex structure as a function of magnetic field and dopant density.  In particular, under and optimal-doped samples may have very different structures from that of the over-doped ones.

The effects of PDW order on the vortex and the associated electronic structures have also been examined in several recent theoretical studies~\cite{Agterberg15b,Dai18,YWang18}. In these approaches, the order parameter fields of a vortex are obtained from phenomenological Ginzburg-Landau free energy functionals. A quadratic lattice fermion model with input from the order-parameter solutions is diagonalized to study the electronic properties, e.g. local density of states, of the vortex. Such semi-empirical approaches can provide qualitatively experimental signatures due to different ordering scenarios, such as whether the checkerboard pattens is driven by a primary PDW order, or the other way around. Yet, a fully self-consistent theoretical modeling of vortex halos, especially one based on well defined microscopic models, is still lacking. Moreover, electron correlation effects, which are known to be important for cuprate materials, cannot be properly included in the phenomenological approaches. 

In this paper, we present a comprehensive study on the structure of vortex halos based on large-scale self-consistent calculations of the well-studied $t$-$J$ model with an additional frustrated second-neighbor $t'$ hopping. It is worth noting that  several advanced many-body techniques, ranging from unrestricted mean-field method and variational Monte Carlo (VMC) simulation to density matrix renormalization group (DMRG) and tensor-network methods, have been applied to study the intertwined orders in strongly correlated electron systems~\cite{Loder11,Himeda02,Berg10,Dodaro17,Jiang18,Zheng17,Corboz14}. In particular, PDW order has been demonstrated in the Hubbard and $t$-$J$ models, as well as their variants. While sophisticated methods, such as DMRG or infinite projected entangled paired states (iPEPS), generally can give more accurate results, and numerically exact results for 1D models, so far they can only be applied to relatively small 2D systems, and hence are not feasible for studying large-scale inhomogeneous states such as superconducting vortices. 

On the other hand, renormalized mean-field theory (RMFT) has proven an efficient method for solving complex structures in $t$-$J$ type correlated electron models~\cite{Garg08,Yang09,Christensen11,Tu16,Banerjee18}. As in most mean-field type approaches, the many-body Hamiltonian is reduced to an effective single-particle problem which is to be solved self-consistently. Yet, unlike standard Hartree-Fock type approximations, the strong electron correlation effects in such systems can be properly captured by the Gutzwiller approach employed in the RMFT~\cite{Gutzwiller1963,Himeda1999,Ogata03}. Importantly, fairly quantitative agreement with STM and ARPES experiments on cuprates have been obtained from RMFT calculations of the frustrated $t$-$t'$-$J$ model~\cite{Choubey17,Choubey20,Choubey21,Tu19,Tsuchiura03}. 
The efficiency of our RMFT calculations is further improved significantly by incorporating the kernel polynomial method (KPM)~\cite{ZWang18,Weisse06} for solving the renormalized tight-binding Hamiltonians, which has to be repeated up to thousand of times in the self-consistency calculation. By applying the RMFT method to the the $t$-$t'$-$J$ model on square-lattice of up to $10^4$ sites, we obtain detailed vortex solutions where up to $10^5$ variational parameters are solved self-consistently. 


Our large-scale RMFT calculations find several vortex solutions with nearly degenerate energies. This result again underscores the complex nature of the superconducting phase, and is reminiscent of previous works showing multiple density-wave states whose energies are almost degenerate with that of uniform $d$-wave SC state~\cite{Tu16,Choubey17,Zheng17,Corboz14}.  In addition to conventional vortices in a uniform $d$-wave background, of particular interest is a self-consistent vortex solution where a bidirectional PDW with a period $8a$ coexists with a charge-density modulation of the same period, which is twice the period of the charge-density wave order resulting from the second harmonic of a unidirectional PDW in the absence of a vortex. This solution, which is obtained without invoking special features of the Fermi surface, is consistent with recent STM experiment~\cite{Edkins19}. 

By carrying out systematic study of the field and doping dependences of these vortex solutions, a coherent and consistent picture is provided for recent STM experiments and the issues of ZBCP. We show that, due to the strong correlation and the orbital $d_{x^2-y^2}$ symmetry, the conductance spectra of STM have sub-gap structures instead of a ZBCP in the under- and optimal-doped regimes even for the plain vortex state. Moreover, the vortex solution with a bipartite checkerboard patterns is shown to result from coexisting bi-directional PDW, CDW, and SDW orders. Such vortices with intertwined orders exhibit energy independent wave vectors at higher energy but with QPI dispersion at low energy. It also shows a larger $s$+$s'$ form factor than that of the $d$-wave symmetry. Finally, the mechanism for the emergence of the checkerboard state inside a vortex is discussed.

\section{Model and methods}

The hole-doped CuO$_2$ plane of copper oxides can be well described by the generalized $t$-$J$ model, which corresponds to the strong coupling limit of the Hubbard model. Its Hamiltonian reads
\begin{eqnarray}
	\label{eq:H_tJ}
	\hat{H} = -\sum_{ij} \sum_{\sigma = \uparrow,\downarrow} \hat{P}_G \, t_{ij} \hat{c}^\dagger_{i\sigma} c^{\,}_{j \sigma}  \hat{P}_G 
	+  J \sum_{\langle ij \rangle} \hat{\mathbf S}_i \cdot \hat{\mathbf S}_j, \qquad
\end{eqnarray}
where $\hat{c}^\dagger_{i\sigma}$ ($\hat{c}^{\,}_{i\sigma}$) is the creation (annihilation) operator of an electron at site-$i$ with spin $\sigma$, and $\hat{\mathbf S}_i = \hat{c}^\dagger_{i\sigma} \mathbf \sigma_{\sigma, \sigma'} \hat{c}^{\,}_{i \sigma'}$ is the electron spin operator at site-$i$,  $t_{ij}$ denotes the electron transfer integral between the Cu $d_{x^2 - y^2}$ orbitals at sites $i$ and $j$. In the $t$-$t'$-$J$ model to be studied in this work, only electron hopping $t$ between nearest-neighbor and $t'$ between next-nearest-neighbor pairs are included in the first term. The notation $\langle ij \rangle$ in the second term denotes the nearest neighbors. The effect of the strong on-site Coulomb repulsion $U \to \infty$ is accounted for by the Gutzwiller projector $\hat{P}_G = \prod_{i} \left(1 - \hat{n}_{i \uparrow} \hat{n}_{i \downarrow} \right)$, which eliminates all doubly-occupied orbitals; here $\hat{n}_{i \sigma} = \hat{c}^\dagger_{i\sigma} \hat{c}^{\,}_{i\sigma}$ is the electron number operator at site-$i$. The residual charge fluctuations between nearest-neighboring pairs $\langle ij \rangle$ leads to the antiferromagnetic superexchange $J \sim t^2/ U$ in the second term above. 
In the presence of a magnetic field $\mathbf B$, its effect is accounted for by the Peierls substitution $t_{ij} = te^{i\phi_{ij}}$, where $\phi_{ij} = \frac{-\pi}{\Phi_0} \int_{\mathbf{r}_i}^{\mathbf{r}_j} {\mathbf A}(\mathbf{r}_i)\cdot d\mathbf{r}_i $, where $\Phi_0=\frac{hc}{2e}$ is the superconducting flux quantum, and $\mathbf A$ is the corresponding vector potential. In our implementation, a Landau gauge $\mathbf A = B (0, x)$ is used.

For large-scale calculations of inhomogeneous states such as those with vortices, the above $t$-$t'$-$J$ model is solved using the RMFT method~\cite{Garg08,Yang09,Christensen11,Tu16,Banerjee18,Choubey17,Tu19,Choubey20,Choubey21,Tsuchiura03}. In this approach, the strong electron correlation, as encapsulated by the projector $\hat{P}_G$, is treated by the Gutzwiller approximation, giving rise to the following renormalized Hamiltonian
\begin{eqnarray}
	& & \hat H_{\rm renorm}=-\sum_{ij,\sigma} g^t_{ij\sigma} \, t_{ij} \left(\hat{c}^\dagger_{i\sigma} \hat{c}_{j\sigma} +  \mbox{h.c.} \right) \\
	& & \qquad +\sum_{\langle i,j\rangle}J\Bigg [ g^{s,z}_{ij} \hat{S}^{z}_i \hat{S}^{z}_j+g^{s,xy}_{ij}\Bigg(\frac{\hat{S}^{+}_i \hat{S}^{-}_j + \hat{S}^{-}_i \hat{S}^{+}_j}{2}\Bigg)\Bigg]. \nonumber
	\label{eq:H_renorm}
\end{eqnarray}
Here $g^t_{ij}$ and $g^{s}_{ij}$ are Gutzwiller factors that account for the renormalization of the hopping amplitude and the exchange interaction, respectively. The exchange interaction $J$, which leads to the SC pairing~\cite{OMahony22} and other intertwined orders, is treated by the Hartree-Fock-Bogoliubov mean-field decoupling. 
In this work, we consider the following mean-field order-parameters: (i) the on-site hole density
\begin{eqnarray}
	\label{eq:delta}
	\delta^v_i = 1 - \langle (\hat{n}_{i \uparrow} + \hat{n}_{i \downarrow} )\rangle,
\end{eqnarray}
(ii) the local magnetization
\begin{eqnarray}
	\label{eq:mz}
	m^v_i = \langle \hat{S}_z \rangle = \frac{\hbar}{2} \langle (\hat{n}_{i \uparrow} - \hat{n}_{i \downarrow} ) \rangle,
\end{eqnarray}
(ii) the bond-order between neighboring Cu sites
\begin{eqnarray}
	\label{eq:chi}
	\chi^v_{ij\sigma} = \langle \hat{c}^\dagger_{i\sigma} \hat{c}^{\,}_{j \sigma} \rangle,
\end{eqnarray}
and (iii) the local electron pairing field on nearest-neighbor bonds ($\bar{\sigma} = -\sigma$)
\begin{eqnarray}
	\label{eq:pairing}
	\Delta^v_{ij\sigma} = \sigma \langle \hat{c}^{\,}_{i\sigma} \hat{c}^{\,}_{j \bar{\sigma}} \rangle.
\end{eqnarray}
Here the superscript $v$ is a reminder that these parameters, while each related to physical observables, are not directly measurable quantities; a proper renormalization factors have to be included for experimental comparisons. The avereage $\langle \hat{O} \rangle = \langle \Psi_0 | \hat{O} |\Psi_0 \rangle$ denotes expectation value of operator $\hat{O}$ with respect to a variational Slater wave function $|\Psi_0\rangle$. The Slater state is constructed from eigenstates of an effective tight-binding Bogoliubov-de Gennes (TB-BdG) Hamiltonian
\begin{eqnarray}
	\label{eq:H_eff}
	& & \hat{H}_{\rm eff} = \sum_{(ij), \sigma} \left( \mathcal{T}_{ij, \sigma} \hat{c}^\dagger_{i \sigma} \hat{c}_{j \sigma} + \mbox{h.c.} \right)
	- \sum_{i,\sigma} \mu_{i\sigma} \hat{n}_{i\sigma} \nonumber \\
	& & \qquad \quad + \sum_{\langle ij \rangle, \sigma} \left( \sigma \mathcal{D}_{ij\sigma} \hat{c}^{\dagger}_{i\bar{\sigma}} \hat{c}^{\dagger}_{j{\sigma}} + \mbox{h.c.} \right) 
\end{eqnarray}
where  $\mathcal{T}_{ij\sigma}$, $\mathcal{D}_{ij\sigma}$, and $\mu_{i\sigma}$ are effective hopping, pairing, and on-site energy parameters, respectively, defined as the derivatives of the variational energy of the renormalized Hamiltonian with respect to the local order parameters:
\begin{eqnarray}
	\label{eq:model_param}
	\mathcal{T}_{ij\sigma} = \frac{\partial W}{\partial \chi^v_{ij\sigma}}, \quad
	\mathcal{D}_{ij\sigma} = \frac{\partial W}{\partial \Delta^{v*}_{ij\sigma}}, \quad
	\mu_{i\sigma} = \frac{\partial W}{\partial n_{i\sigma}}.
\end{eqnarray}
Here $W = \langle \Psi_0 | \hat{H}_{\rm renorm} | \Psi_0 \rangle$ supplemented by Lagrangian multipliers for enforcing the wave function normalization and the conservation of electron number, and $n_{i\sigma} = \langle \hat{n}_{i\sigma} \rangle $ is the local electron number; see Appendix~\ref{sec:RMFT} for details. 
These parameters in turn depend the local order-parameters listed in Eqs.~(\ref{eq:delta})--(\ref{eq:pairing}), that are determined self-consistently. 
In order to allow for spatial inhomogeneity, these site and bond-dependent parameters, with a total number of the order of $10^5$, are treated as independent and optimized in real-space RMFT calculations; details of the method are presented in Appendix~\ref{sec:RMFT}.

In the RMFT calculations, the numerous local mean-field parameters are solved self-consistently through iterations. For systems with up to $N = 10^4 \sim 10^5$ sites, each iteration requires solving a large $2N \times 2N$ tight-binding matrix. For unrestricted optimizations, a number of $1000\sim 3000$ iterations are routinely required to reach satisfactory convergence. Although the TB-BdG Hamiltonian can be exactly solved by the exact diagonalization (ED), the poor $\mathcal{O}(N^3)$ scaling of ED renders it infeasible for large-scale calculations. Here, instead, we incorporate the kernel polynomial method (KPM)~\cite{Weisse06,Barros13,ZWang18}, which provides a linear-scaling $\mathcal{O}(N)$ approach to electronic structure problems, into the RMFT framework.

In this approach, the total or local DOS of the system is expanded as a Chebyshev polynomial series whose coefficients are efficiently computed through matrix-vector products, which can be made linear-scaling for large sparse TB matrices. The various correlation functions $\langle c^\dagger_{i\sigma} c^{\,}_{j \sigma'} \rangle$ and $\langle c^{\,}_{i \sigma} c^{\,}_{j \sigma'} \rangle$ required for the computation of local order parameters are obtained through automatic differentiation. We note in passing that KPM has been used within the conventional Hartree-Fock-Bogoliubov mean-field to study large-scale structures of SC states~\cite{Covaci10,Nagai12,Berthod16}. Yet, the introduction of the Gutzwiller renormalization factors significantly increases the computational complexity of RMFT. In our implementation, the efficiency of KPM is further enhanced by employing the probing method~\cite{Silver94,Tang12} for estimating the trace of large matrices and the utilization of GPU programming; more details can be found in Appendix~\ref{sec:kpm}.

The KPM-RMFT method discussed above is mainly used to obtain the various local parameters in a vortex structure, a calculation which often requires up to thousands repeated solutions of the TB-BdG Hamiltonians in order to reach self-consistency. Once these local orders are obtained, a hybrid approach is employed to compute the corresponding electronic structures such as local density of states. We combine efficient exact diagonalization (ED) package~\cite{tdb10} with the supercell method~\cite{Zhu02,Schmid10} to obtain the exact eigenfunctions of the converged TB-BdG equation. We note that even though large-scale ED is time-consuming, this is a one-shot calculation using the converged order-parameters. The supercell method allows us to further reduce the finite-size effect and enhance the momentum resolution.

The electron Green's function, which is directly related to several experimental measurements, can be computed from the exact eigenfunctions of the renormalized TB-BdG Hamiltonian. In particular, the local density of states (LDOS) at the $i$-th site corresponds to the diagonal element of the lattice Green's function 
\begin{eqnarray}
	\label{eq:N_i}
	N_i(\omega) =-\frac{1}{\pi} {\rm Im}G_{ii}(\omega).
\end{eqnarray}
For better comparisons with experiments, one can take into account the effects of electron orbitals using the  continuum Green's functions~\cite{Choubey14,Kreisel15} defined as 
\begin{equation}\label{eq:GWW}
	G(\mathbf{r},\mathbf{r^\prime},\omega) =  \sum_{ij} G_{ij}(\omega) W_i(\mathbf{r}) W^*_j(\mathbf{r^\prime}).
\end{equation}
Here $W_i(\mathbf{r})$ is the Wannier function centered at site $i$. The corresponding continuum LDOS, which can be directly measured by the tunneling conductance, is 
\begin{eqnarray}
	\label{eq:g_r}
	g(\mathbf{r},\omega) = -\frac{1}{\pi} {\rm Im} G(\mathbf{r},\mathbf{r},\omega).
\end{eqnarray}	
In practical calculations, a broadening factor $\Gamma$ is used to incorporate the finite lifetime or scattering effect, which effectively replaces the delta function $\delta(\omega \pm E_n)$ of the ${\rm Im}G$ by a Lorentzian function $1/[(\omega \pm E_n)^2 +\Gamma^2]$. In this work, the broadening parameter $\Gamma$ is assumed to have an energy dependent form, $\alpha|\omega|+3.0\times 10^{-4}$, where $\alpha=0.25$ is shown to give best agreement with experiments \cite{Choubey17,Choubey21,Alldredge08}.	
In this work, the parameters for the $t$-$t^\prime$-$J$ model are set to be $t^\prime/t=-0.3$, $J/t=0.3$. For comparison with experiments, we use $t = 400$ meV, which serves as the reference for energies.  The lattice sizes considered are $N_x\times N_y=48\times48$, $48\times96$, and $96\times 96$, which correspond approximately to magnetic fields $B\sim 11.92$, $5.96$ and $2.98$ Tesla, respectively, assuming a lattice constant $a=3.88 \angstrom$.  The number of magnetic supercells $M_x \times M_y$ are selected to ensure that $M_x \times N_x = M_y \times N_y = 384$.

\begin{figure*}[t]
\includegraphics[width=2.0\columnwidth]{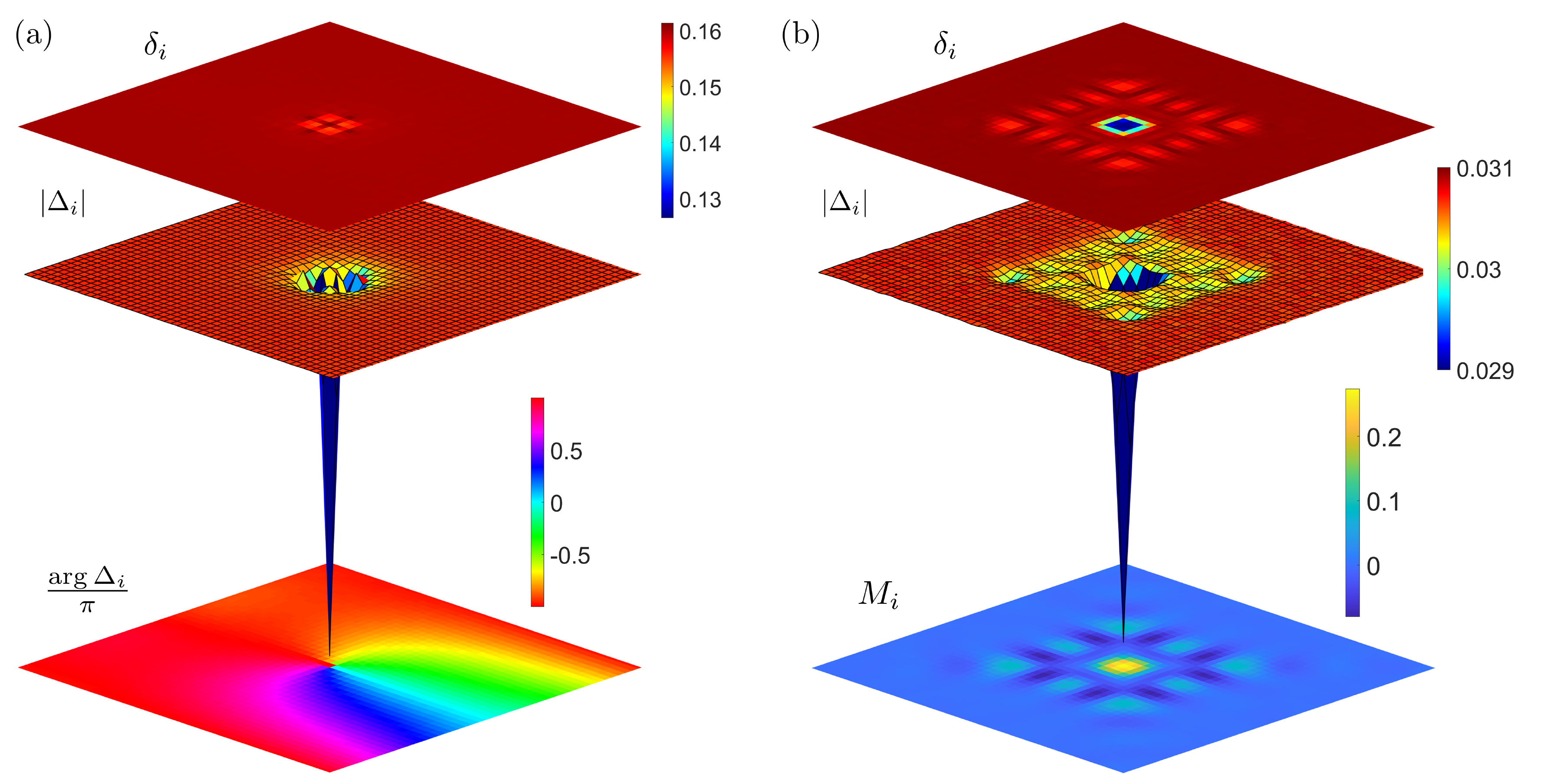}
\caption{ 
	Site-dependent order parameter $\delta_i$, $\Delta_i$, and $M_i$ of simple $d$-wave vortices centered at $(23,23)$.  The dopant concentration is $x=0.16$, and lattice size is $48\times96$ hence the corresponding magnetic-fieled is $B=5.96T$.  
	(a) The plain-vortex solution. (b) The CB-h solution.
	On the top of (a) and (b) shows the 2D map of $\delta_i$.
	In the middle of (a) and (b) shows the 3D map of pairing amplitude $|\Delta_i|$.
	The bottom of (a) shows the pairing phase $\arg \Delta_i/\pi$.
	The bottom of (b) shows the 2D map of staggered magnetization $M_i$.
}
\label{fig:0}  \
\end{figure*}
	
\section{Results}	

\label{sec:results}

To obtain vortex solutions in the RMFT calculation, the various local order parameters need to be properly initialized. For example, a point singularity or $2\pi$ vorticity has to be introduced to the phase of the pairing parameters $\Delta_{ij}$. Similarly, proper modulations of the pairing amplitudes, bond variables, and hole densities are required for the initial state of the RMFT iterations. Depending on the initial conditions, our calculations uncover several vortex solutions with nearly degenerate energies, including, among others, plain $d$-wave vortex and one with intertwined PDW/CDW orders. The many vortex solutions with comparable energies obtained in our studies are reminiscent of the multitude of bulk states with intertwined orders that compete with the uniform $d$-wave SC state~\cite{Tu16}. These intriguing results from the generalized $t$-$J$ models, including both bulk and vortex solutions, highlight the complex nature of cuprate superconductors and the importance of extrinsic effects such as disorder and impurities. 

In this work we consider two particular vortex solutions, which are most relevant to experiments. The first one, to be called plain vortex in the following, can be viewed as the natural topological defects of the uniform $d$-wave SC state. The corresponding pairing field is not accompanied by either PDW or CDW orders. The second solution describes a vortex structure with a checkerboard modulations in both pairing, charge-density, and magnetization. The structures of these two vortex solutions on a $N_x \times N_y = 48\times 96$ square lattice are shown in Fig.~\ref{fig:0}(a) and (b). An average hole density $\delta = 0.16$ and a magnetic field $B = 5.96$~T is used in the KPM-RMFT calculation. Because of the periodic boundary conditions, a configuration with two vortices separated by half the linear size in the $y$ direction was used as the initial state. Large-scale calculations are necessary here to ensure that $N_y, N_y \gg \xi$, where $\xi$ is the characteristic size of vortex, hence minimizes the finite size effect.

\begin{figure*}
\includegraphics[width=2.0\columnwidth]{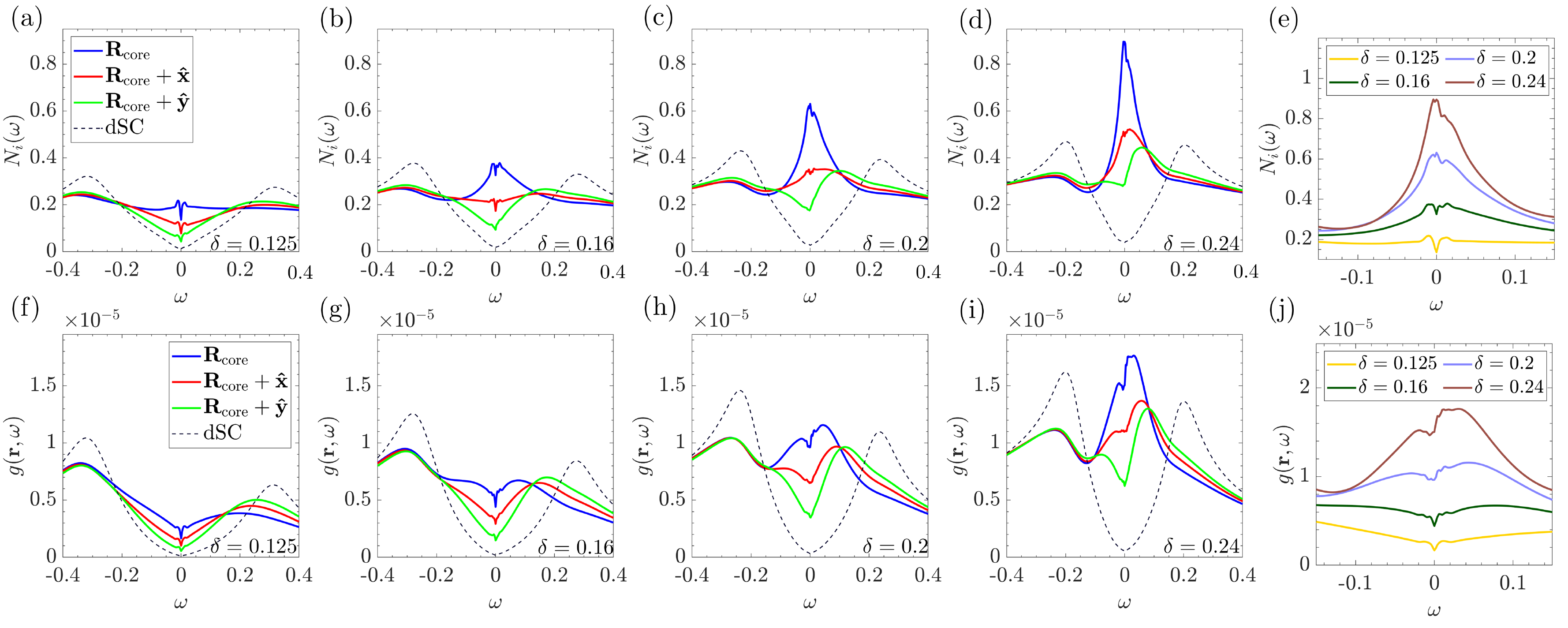}
\caption{ Evolution of the electronic structure near the vortex core for plain-vortex state at $B=5.96$T. Panels~(a)-(e) is the LDOS $N_i(\omega)$ while the corresponding continuum LDOS $g(\mathbf r,\omega)$ are shown in panels~(f)-(j). The dopant in the (a)-(d) and (f)-(j) are $\delta=0.125$, $\delta=0.16$, $\delta=0.2$, and $\delta=0.24$. Panels~(e) and (j) plot the low-energy spectrum of the discrete and continuum LDOS, respectively, at the vortex core for four dopant data together for comparison. }
\label{fig:B6Tall}  \
\end{figure*}

At the top of Fig.~\ref{fig:0}(a) and (b) are the density plots of the on-site hole density $\delta_i$ around the vortex for plain-vortex and checkerboard-halo solution, respectively.
For both types of vortices, the suppression of the pairing field at the vortex core leads to an excess hole density $\delta_{\rm core}$. Importantly, the hole density $\delta_i$ of the second vortex solution exhibits a checkerboard halo extending over tens of lattice constants. 
Next we examine the configuration of the pairing field. To that end, we first define a physical pairing order parameter, which takes into account the local $d_{x^2-y^2}$ structure and the renormalization effect: 
 \begin{equation}\label{eq:giDelta}
 	\Delta_i= \frac{1}{8}\sum_{\sigma,\mathbf{\tau}}g^t_{i,i+\mathbf{\tau},\sigma}(-1)^{\tau_y}\Delta_{i,i+\mathbf{\tau},\sigma} \, e^{i\frac{\pi}{\Phi_0} \int_{\mathbf{r}_i}^{\mathbf{r}_j} {\mathbf A}(\mathbf{r}_i)\cdot d\mathbf{r}_i},
 \end{equation}
where $\mathbf \tau$ represents the unit vectors $\pm \hat{\mathbf x}$ or $\pm \hat{\mathbf y}$.  The corresponding 3D map of the pairing amplitude $\Delta_i$ is shown in the middle of Fig.~\ref{fig:0}(a) and (b) for the two different vortex solutions. The point singularity of the pairing field is demonstrated in the bottom panel of Fig.~\ref{fig:0}(a) for the case of plain vortex, which shows a clear branch cut with a $2\pi$ phase jump running along the negative $x$ direction.  The amplitude of the pairing field can be well approximated as $|\Delta(\mathbf r_i)| \sim \Delta_0 \, \tanh(|\mathbf r_i| / \xi_c)$, where the vortex size is estimated to be $\xi_c \sim 3a$ from the fitting. Using a lattice constant $a=3.88${\AA} for the Cu-O plane, this corresponds to a vortex size of $2 \sim 3$~nm, which agrees well with the experimental result~\cite{Edkins19}. Finally, the bottom panel of Fig.~\ref{fig:0}(b) illustrates the staggered magnetization $M_i$ of the checkerboard-halo solution. Taking into account the renormalization effect, the staggered magnetization is given by
 \begin{equation}\label{eq:gm}
	M_i=(-1)^{x+y}  \frac{g^{s,xy}_i}{4} \sum_{\mathbf \tau} \sqrt{\frac{g^{s,z}_{i,i+\mathbf{\tau}}}{g^{s,xy}_{i,i+\mathbf\tau}} } m_i,
\end{equation}
where $\mathbf\tau$ represents the unit vectors $\pm \hat{\mathbf x}$ and $\pm \hat{\mathbf y}$, and $g^{s, xy}$, $g^{s, z}$ are the Gutzwiller factors for the exchange interactions~\cite{Mi}.
Below we present details of these vortex structures and their experimental manifestations.

\subsection{Tunneling conductance} 


The tunneling conductance measured by STM is related to the LDOS defined in Eqs.~(\ref{eq:N_i}) and its continuous version (\ref{eq:g_r}), which takes into account the structure of the electron orbitals through the Wannier functions. For comparison, we first show in Figs.~\ref{fig:B6Tall}(a)-(d) the bare LDOS,~$N_i(\omega)$, at the center $\mathbf R_{\rm core}$ of the vortex under a magnetic field of $5.96$ T for four different dopant densities $\delta=0.125$, $0.16$, $0.2$ and $0.24$. Also shown for comparison is a reference LDOS corresponding to the bulk of the uniform $d$-wave SC. The spectra at the vortex core exhibit a pronounced peak at the Fermi level, especially for large hole doping as shown in Fig.~\ref{fig:B6Tall}(b), (c), and (d). Interestingly, this central peak of the LDOS is immediately suppressed even at locations one lattice constant away from the center. 
For better comparison, the LDOS in the vicinity of the Fermi level is shown in Fig.~\ref{fig:B6Tall}(e) for the four different dopant densities. The large enhancement near zero energy increases rapidly with doping.

For the underdoped (UD) case with $\delta=0.125$, as shown in Fig.~\ref{fig:B6Tall}(a), the enhancement of the LDOS is relatively small compared with those at larger dopings. Moreover, a subgap structure develops for very small energy $|\omega| \lesssim 0.01$. This subgap structure persists at optimally doped (OP) $\delta=0.16$ [Fig.~\ref{fig:B6Tall}(b)] and becomes invisible in the overdoped (OD) regions of $\delta=0.2$ and $0.24$ [Fig.~\ref{fig:B6Tall}(c) and (d)]. The general features of OD results are in good agreement with calculations in Refs. \cite{Wang1995, Franz1998}. This is expected since those calculations have not considered the strong coupling renormalization effect, which is less important in OD.

In order to compare with the tunneling conductance in the vicinity of a SC vortex reported in recent STM experiments \cite{Renner21, Machida16, Edkins19}, Figs.~\ref{fig:B6Tall}(f)-(i) show the computed continuum LDOS $g(\mathbf{r},\omega)$ for the same four dopant densities inside a vortex of $5.96T$ field. Compared with the bare spectra $N_i(\omega)$, the continuum LDOS is suppressed at low energies at the vortex core, especially for the UD $\delta$=0.125 and OP $\delta$=0.16 cases. Indeed, the conductance in both low-doping cases is expected to exhibit a suppression at low energies, in stark contrast to the enhanced conductance of the lattice LDOS $N_i(\omega)$. On the other hand, a weak enhancement or ZBCP at low energies remains only in the OD cases of $\delta=0.2$ and $0.24$, as shown in panels~(h) and (i) of Fig.~\ref{fig:B6Tall}.
 Finally, we note that the conductance at the nearest-neighbor sites of the core, the red and green curves in Figs. \ref{fig:B6Tall}(f)-(i) is suppressed at low energies for all dopant densities.

Importantly, our results resolve a long-standing puzzle concerning the issue of ZBCP. Despite the prediction of its existence by Wang and MacDonald~\cite{Wang1995}, the ZBCP was not detected in almost all of the early STM measurements~\cite{Machida16, Edkins19}. The absence of the ZBCP, in our view, can be attributed to the fact that these early experiments are mostly carried in either the UD or OP samples which are expected to exhibit a suppression of conductance near zero energy according to our calculations shown in Figs.~\ref{fig:B6Tall}(f) and (g). On the other hand, our calculation shows that the ZBCP remains in the case of large doping, as illustrated in Fig.~\ref{fig:B6Tall}(h) and (i). Such enhanced conductance at $\omega \sim 0$ was indeed observed in the OD sample of BSCO~\cite{Renner21}.

The qualitatively different behaviors of the tunneling conductance between the UD and OP cuprates and the OD ones are mainly due to two reasons. First, as demonstrated by Wang and MacDonald in their conventional mean-field calculation of a $d$-wave vortex~\cite{Wang1995}, the enhancement of the LDOS can be attributed to the resonant states in the vortex core. Such resonance, however, is suppressed due to strong electron correlation at small hole densities in the UD and OP cuprates. Since the renormalization effect is less significant at large hole density, a resonance-induced conductance remains in the OD samples. The second reason is that the STM measurement involves the contribution from the Wannier function used in Eq. \eqref{eq:GWW}, which has copper  $d_{x^2-y^2}$  orbital symmetry. This symmetry strongly suppresses the contribution at the copper site, especially along the $y=\pm x$ directions, as shown in Fig. 2 in Ref.~\cite{Choubey17b}. 
Thus, the ZBCP vanishes and a sub-gap-like structure emerges in the UD and OP samples. 
ZBCP in the OD samples is also greatly reduced but still visible.

Another interesting observation mentioned by Gazdi\ifmmode \acute{c}\else \'{c}\fi{} \etal~\cite{Renner21} is that the interaction between vortices or vortices at higher field could change details of the conductance spectra near zero energy. In the SM, the LDOS of the lattice model inside a vortex under   $2.98T$, $5.96T$ and $11.92T$ magnetic field is shown in Fig.~S4 for the same four dopant densities as Fig.~\ref{fig:B6Tall}. The distances between two vortices are $24\sqrt{2}$, $48$ and $48\sqrt{2}$ and the corresponding fields are $B=11.92T$, $5.96T$ and $2.98T$, respectively. Within this range of distances we have not found significant differences in spectra
except that the sub-gap structure of the tunneling conductance at high field 11.92T becomes more pronounced. This is in qualitative agreement with experiment \cite{Renner21}.
The second issue raised in Ref. \cite{Renner21} about the QPI  versus charge order will be discussed later after we discuss the checkerboard-halo state.

\begin{figure*}
\includegraphics[width=1.85\columnwidth]{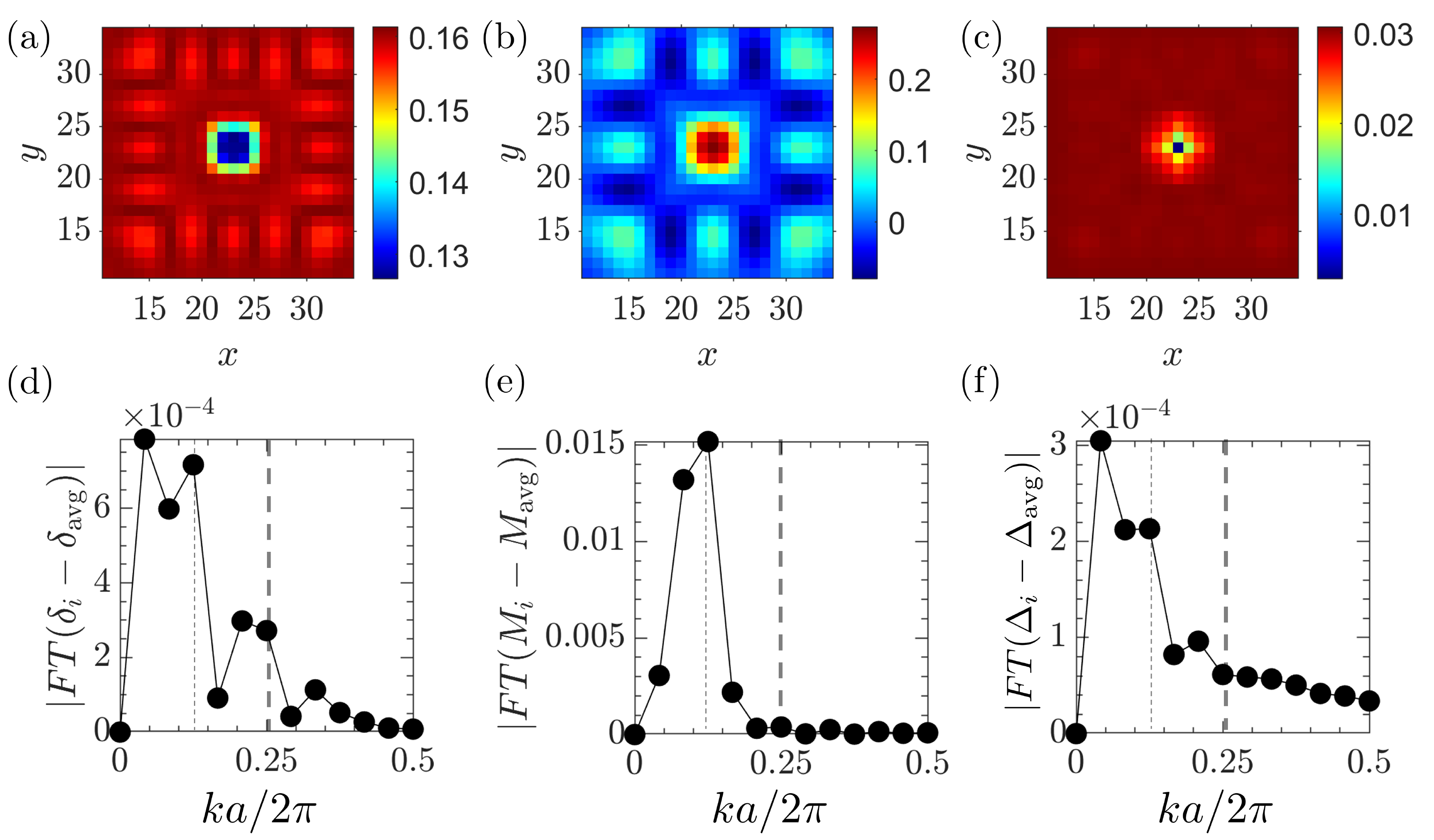}
\caption{Spatial profile of the intertwined order parameters of the CB-h vortex solution at $B = 5.96T$ and $\delta=0.16$: (a)~hole density $\delta_i$, (b) staggered magnetization $M_i$ and (c) pairing amplitude $|\Delta_i|$. The corresponding discrete Fourier transform of the spatial pattern along the $k_x$ axis are shown in panels~(d), (e) and (f), respectively. The think and thick dashed lines indicate the characteristic wave vectors $\mathbf Q_x = (1/8, 0) (2\pi/a)$ and $2\mathbf Q_x = (1/4, 0) (2\pi/a)$, respectively.}
\label{fig:sCBhalocfg}  \
\end{figure*}


\subsection{Checkerboard Halo state}

So far, the result of the plain-vortex could explain the experiment at low field and in the OD regime of cuprates.
However, recent STM experiment~\cite{Edkins19} found a bi-directional PDW state inside a vortex halo. This PDW-vortex mixed state exhibits CDW peaks at $\mathbf{Q}_x=(2\pi/8a,0)$  and $2\mathbf{Q}_x=(2\pi/4a,0)$ and in $y$-direction, as well. It is also worth noting that the CDW peaks are energy-independent~\cite{Machida16}, which is inconsistent with  the plain-vortex solution. 
In addition, experiments found a dominant  $s$+$s^\prime$-like form factors in the conductance map. All these features can be consistently explained based on a self-consistent vortex solution in our large-scale RMFT calculations. As this vortex solution exhibits a bi-directional PDW state inside a vortex halo, it will be called the checkerboard-halo (CB-h) solution.

The spatial profiles of the three order parameters, hole density $\delta_i$, staggered magnetization $M_i$, and pairing amplitude $\Delta_i$, of the CB-h state obtained assuming a dopant density $\delta=0.16$ and a magnetic field $B=5.96T$, are shown in Figs.~\ref{fig:sCBhalocfg}(a), (b), and (c) respectively. The calculation was performed on a $48\times 96$ lattice which contains two vortices separated by a distance of 48 lattice constants. The size of the halo is about $24\times 24$ around the vortex center.  Outside the halo, all order parameters quickly relax to the values of a uniform $d$-wave pairing state in the absence of a magnetic field. Only staggered magnetization still has a vary small value about $0.02$ left outside the halo. Figs.~\ref{fig:sCBhalocfg}(d)-(f) show the discrete Fourier transform of the hole density, staggered magnetization, and pairing amplitude, respectively. In order to highlight the nature of the spatial modulations, the constant background values of these order-parameters are removed; these background values are listed in Table.~SII of the supplemental material. As the CB-h vortex solution preserves the tetragonal symmetry, the profiles of these order parameters are the same along $x$ and $y$ directions.

The Fourier transforms of all three order parameters exhibit a clear peak at the wave vector $\mathbf Q_x$, while the peak at $2\mathbf Q_x$ is more evident for the charge density modulation. On the other hand, the $2\mathbf{Q}_x$ peak is essentially negligible for staggered magnetization.   It is worth noting that such CDW order with both wave vectors $\mathbf Q$ and $2\mathbf Q$ agrees well with recent STM experiment~\cite{Edkins19}. We also note that both hole density and pairing amplitude shows a peak at small $k_x$, which is simply due to the smooth variation of the vortex profile, and is unrelated to the modulation. A similar peak is observed in the discrete Fourier transform of a plain vortex state, as shown in Fig.~S2 in the supplemental material.

The structure of the CB-h vortex at different hole densities as well as the magnetic field effects are also systematically investigated. In general, the $\mathbf Q$-peak of the CDW is more pronounced than the $2\mathbf Q$ peak. Yet, this difference between $\mathbf Q$ and $2\mathbf Q$ is reduced with increasing doping. On the other hand, the $2\mathbf Q$ peak of the CDW is enhanced by the magnetic field relative to that at $\mathbf Q$. Indeed, in the presence of a large magnetic field, the period-$4a$ modulation can become even more prominent than that of the $8a$ period for the OD cases.  More details of these doping and field dependences can be found in Fig.~S8 to S10 in the supplemental material. 

Our calculation also shows that the bi-directional checkerboard of the vortex halo is accompanied by a bi-directional SDW with an $8a$ period in both $x$ and $y$ directions; see Fig.~\ref{fig:sCBhalocfg}(b) and (e).  In the STM experiment in Ref.\cite{Edkins19}, it is unclear whether there is an associated SDW order with their observed PDW order. In an early neutron scattering experiment \cite{Lake01,Lake02}, the field-induced AFM order is reported for the optimally doped La-based cuprates. Similar conclusions have been reached for YBCO \cite{Mitrovic03}. But no similar result is yet to be reported for BSCO.

\begin{figure*}
\includegraphics[width=2.0\columnwidth]{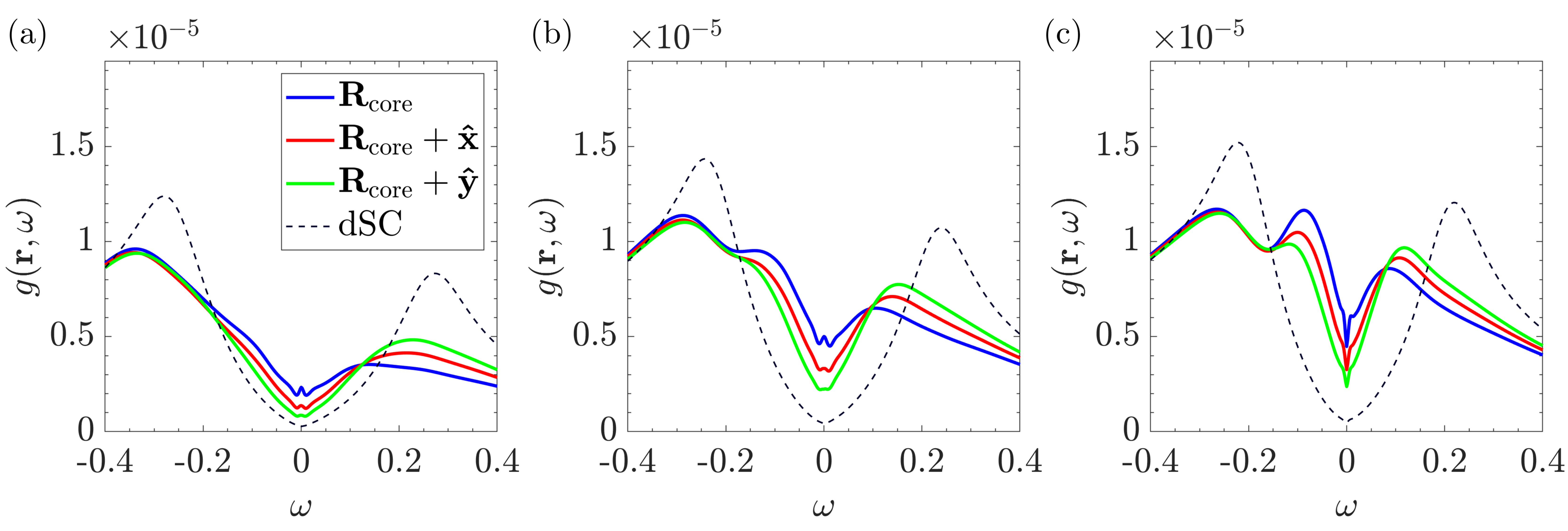}
\caption{Conductance of the CB-h solution at $B=5.96T$ for (a) $\delta=0.16$, (b) $\delta = 0.2$ and $\delta = 0.22$.  The black dashed line is the conductance far from the vortices core, which approach to uniform SC case.}
\label{fig:sCBhalo}  \
\end{figure*}

In Fig.~\ref{fig:sCBhalo}, the tunneling conductance of the CB-h state is shown for dopant densities $\delta=0.16$, $0.2$ and $0.22$, at a magnetic field of $B=5.96T$. The conductance of the uniform $d$-wave SC, shown as the black-dashed line, serves as a reference.  As expected, the energies of the two coherent peaks shrink with increasing dopant density from the OP to the OD regime. At the optimum doping $\delta=0.16$, a tiny peak appears near the zero bias, which evolves into a sub-gap structure, as exemplified by a dip in Fig.~\ref{fig:sCBhalo}(c), as the dopant increases.
In the presence of a magnetic field, the small peak at $\omega \sim 0$ at the OP is also suppressed and disappears at $B=11.92T$, which is shown in the SM Fig.~S7(a).

Most notably, contrary to the case of plain vortices shown in Fig.\ref{fig:B6Tall}, there is no enhancement of conductance at all dopant densities. Indeed, there is little difference in the spectra at the OP doping $\delta=0.16$ between the plain and Ch-B vortices. On the other hand, as discussed previously, the plain-vortex solution exhibits an enhancement of the cLDOS in the large hole density OD regime at the vortex core; see e.g. Fig.\ref{fig:B6Tall}~(i). Yet, the conductance spectra of the CB-h vortex at OD is marked by a sub-gap structure as shown in Fig.~\ref{fig:sCBhalo}(c). This difference in conductance spectra thus could serve as the indicator of the presence of the PDW in the vortex for the OD samples. However, as both plain and CB-h vortices show similar conductance spectra in the UD and OP cases, the spectral detection of the PDW requires a more careful examination of the spatial dependence of the spectra by taking into account the various form factors.

\begin{figure*}
\includegraphics[width=2.0\columnwidth]{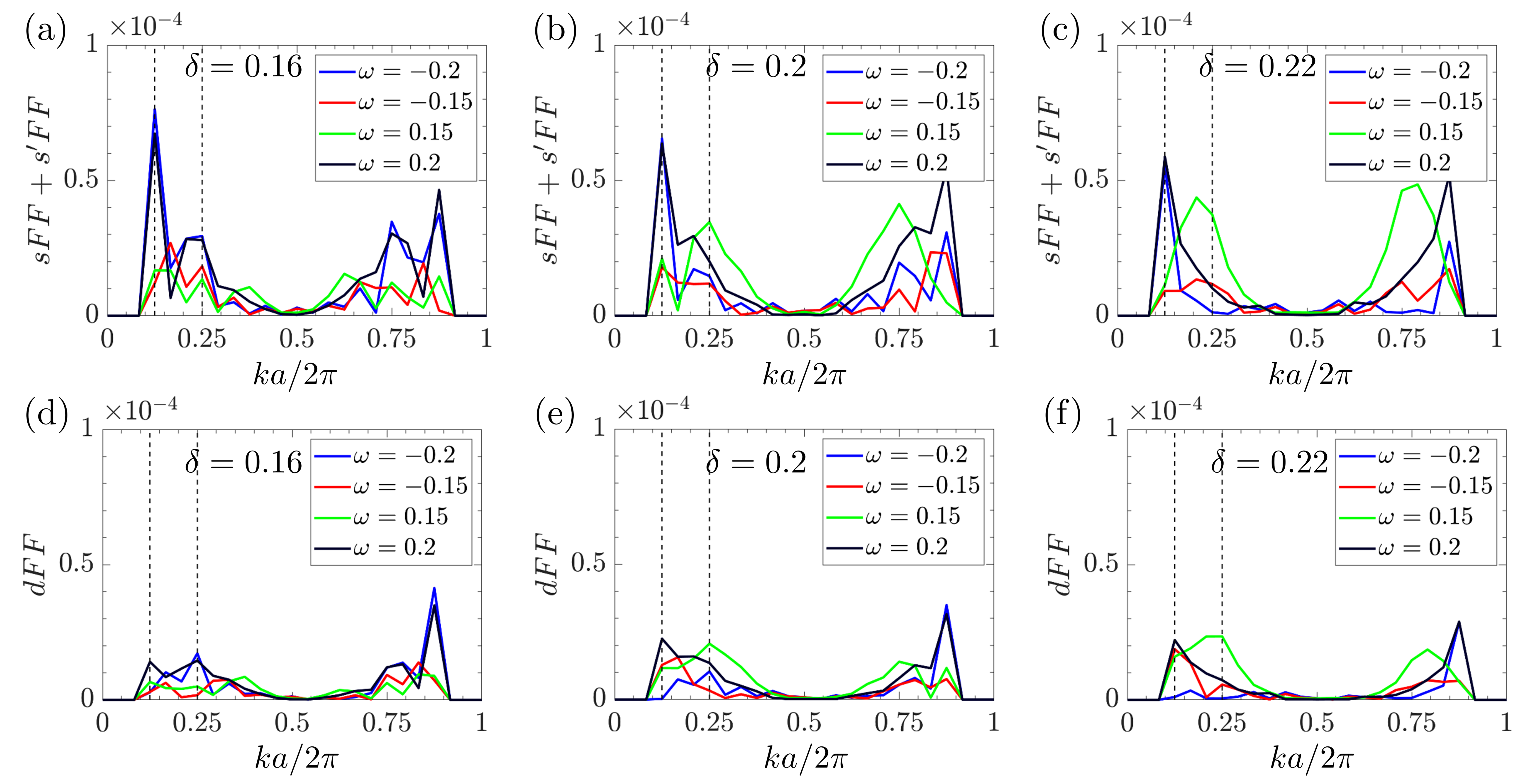}
\caption{Form factors of the the CB-h vortex solution at $B=5.96T$ for four different energies $\omega = \pm 0.15$ and $\pm 0.2$. The three top panels show the $s$+$s^{\prime}$-form factors at dopant densities (a) $\delta=0.16$, (b) $\delta = 0.2$, and (c) $\delta = 0.22$, while the corresponding $d$-form factors are shown in the bottom three panels~(d)--(f). The two dashed lines indicate the characteristic wave vectors $\mathbf Q_x = (1/8, 0) (2\pi/a)$ and $2\mathbf Q_x = (1/4, 0) (2\pi/a)$.}
\label{fig:ffs}  \
\end{figure*}

To this end, the form factors are calculated by first obtaining the sublattice conductance map or $g$-map, which can be separated into contributions from the copper and oxygen sites at distinct nearest-neighbor bonds $Cu(\mathbf{r},\omega)$, $O_x(\mathbf{r},\omega)$ and $O_y(\mathbf{r},\omega)$. In the $t$-$J$ model, the copper sites correspond to the lattice point $\mathbf R$ of the square lattice, which means $Cu(\mathbf{r})=g(\mathbf{R})$. The oxygens sit on the nearest-neighbor $x$ or $y$ bonds, the oxygen conductance thus could be recognized as $O_x(\mathbf{r}) = g(\mathbf{R}+0.5\mathbf{\hat x})$ and $O_y(\mathbf{r}) = g(\mathbf{R}+0.5\mathbf{\hat y})$.
After the Fourier transform of these quantities, one obtains the following form factors:
\begin{equation}
\begin{aligned}
&	s(\mathbf{q},\omega) = Cu(\mathbf{q},\omega),           \\
&	s^\prime(\mathbf{q}, \omega) = \frac{1}{2}\left[ O_x(\mathbf{q},\omega)+O_y(\mathbf{q},\omega) \right],         \\
&	d(\mathbf{q},\omega) = \frac{1}{2} \left[ O_x(\mathbf{q},\omega)-O_y(\mathbf{q},\omega) \right].         \\
\end{aligned}
\end{equation}
The form factors of the CB-h vortex at a magnetic field $B = 5.96$T are summarized in Fig.~\ref{fig:ffs}.  The three top panels (a)--(c) show the $s$+$s'$ form factors at dopant densities $\delta = 0.16$, $0.2$, and 0.22, respectively, while the corresponding $d$ form factors are shown in the bottom three panels (d)--(f). Here we focus on the low energy regime where $|\omega| \lesssim 0.2$. This energy range accounts for about 50\% to 75\% of the coherence peak of $d$-wave pairing state. It is about the same energy range used in experiments of Ref.~\cite{Edkins19} with form factors shown at 30 meV while the coherence peak is about 40 meV. A systematic investigation of the energy dependence can be found in the supplemental material.

\begin{figure*}
	\includegraphics[width=2.0\columnwidth]{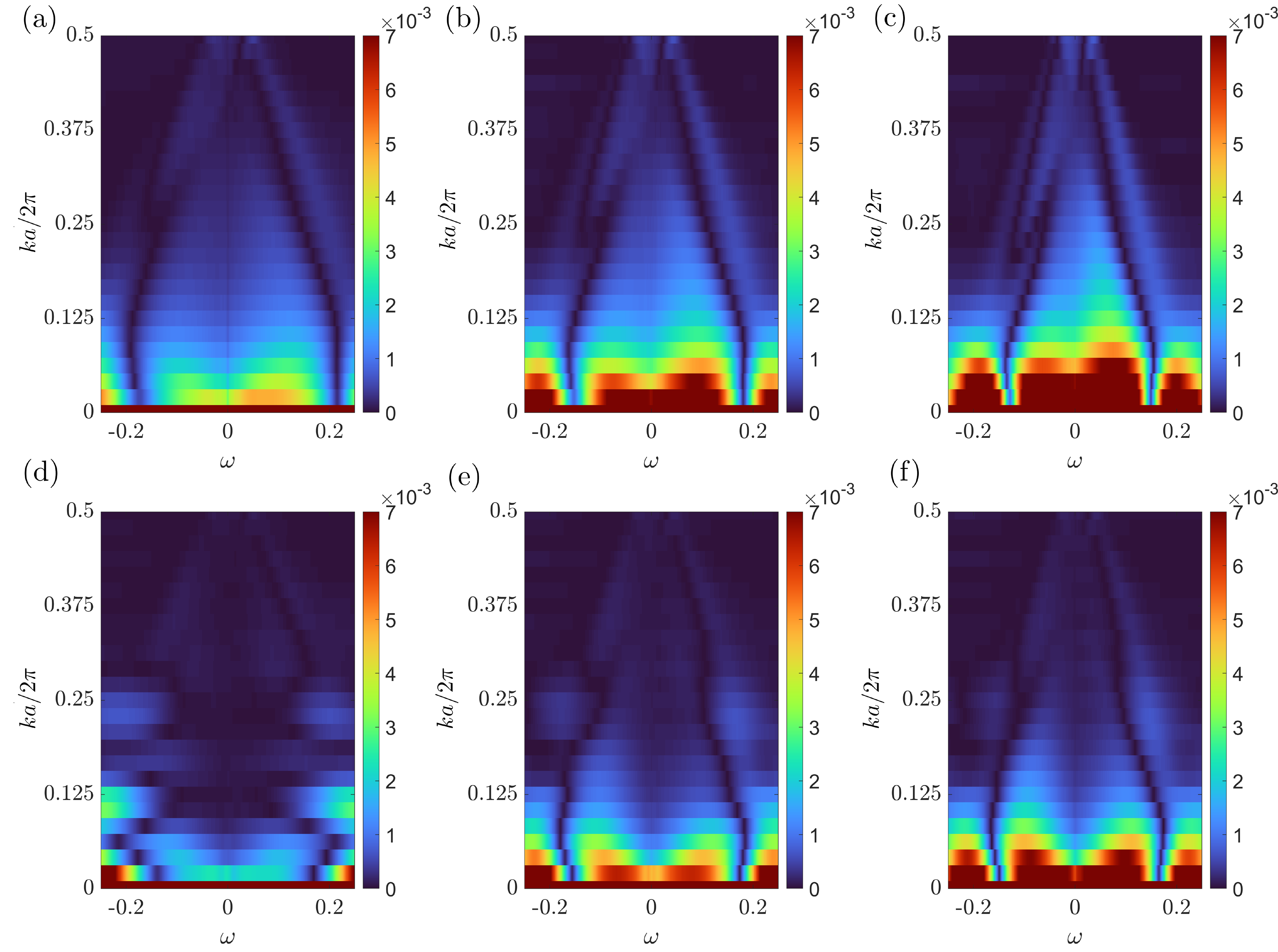}
	\caption{ Electronic structure of a vortex state in the momentum space under a magnetic field $B=2.98T$. The discrete Fourier transform of the conductance $|g(\mathbf{k},\omega) |$ is shown along the $k_x$ direction within the first Brillouin zone. Panels (a)-(c) shows the case for a plain-vortex state at dopant concentration $\delta=0.16$, $\delta=0.2$ and $\delta=0.24$, respectively. The Fourier transform of the conductance maps of the CB-h vortex at these three doping are shown in panels~(d)-(f), respectively.}
	\label{fig:gmapFT_3T}  \
\end{figure*}	

One of the interesting STM results of Ref.~\cite{Edkins19} is the observation of the same form factors at the positive and negative bias of 30 meV. This low-energy particle-hole symmetry of the form factors is also reproduced in our RMFT calculations, as both the $s+s^\prime$ and $d$ form factors are nearly identical for $\omega = \pm 0.2$ and $\pm 0.15$; see Fig.~\ref{fig:ffs}(a) and (d) for $\delta = 0.16$. This dopant density is also close to the hole density 0.17 reported in the same STM experiment~\cite{Edkins19}. Similar particle-hole symmetry also persists in the presence of magnetic field at $B = 2.98$T and 11.92, as demonstrated in the supplemental material. However, the disparity between the positive and negative bias becomes more evident with increasing doping toward the OD regime.

Another important feature of the CB-h vortex is a more substantial $s$+$s'$ form factor than the $d$-form factor, a result which is also consistent with the Fourier analysis of the STM experiment~\cite{Edkins19}.
In our calculations, this dominance of $s$+$s^\prime$ over $d$ form factors can be attributed to the presence of SDW order in addition to CDW and PDW within the vortex halo. Indeed, as pointed out in our previous works on the bulk PDW order, the conductance map is dominated by the $d$-form factor in the absence of the SDW order. One of our important prediction is thus the existence of a SDW coexisting with both PDW and CDW in the vortex halo state reported in the recent STM experiment~\cite{Edkins19}.

\subsection{Bipartite Vortex Structure between QPI and Charge Order }

Another intriguing puzzle related to the checkerboard patterns in a vortex halo is the role played by the quasi-particle interference (QPI). For example, a recent experiment~\cite{Renner21} on heavily over-doped Bi2212 sample reveals a checkerboard in the vortex halo under a magnetic field $B = 3$T.  Interestingly, for energy less than half of the superconducting gap, the peaks at the characteristic wave vectors $2\mathbf{Q}_{x} =(\pm \frac{2\pi}{4a},0)$ and $2\mathbf{Q}_y = (0,\pm \frac{2\pi}{4a})$ of the $g(\mathbf k,\omega)$ map seem to be energy dependent. A similar energy-dependent bipartite structure has also been reported on an OP Bi2212 sample at $B = 11.25$T by Machida {\em et al.}~\cite{Machida16}, who suggested that these period-$4a$ modulations could arise from an enhanced QPI at low energy, and become energy independent at larger $\omega$.  

To investigate this scenario, we examine the Fourier transform $g(\mathbf k,\omega)$ of the conductance map as well as the QPI signatures for both a plain vortex and a CB-h vortex at a magnetic field $B = 2.98$T.  First, the results of a plain vortex are shown in Figs.~\ref{fig:gmapFT_3T}(a)-(c) for three hole densities $\delta=0.16$, $0.2$, and $0.24$, respectively. The overall conductance behaviors are rather similar for the three dopant densities. The conductance is dominated by contributions from small wave vectors. Moreover, as expected, there are no special features at the characteristic wave vectors $ka/2\pi =0.125$ ($\mathbf{Q}$) and $0.25$ ($2\mathbf{Q}$) for such a plain vortex. 

The conductance map of a CB-h vortex, on the other hand, exhibits a more complex energy-momentum dependence, as demonstrated in Figs.~\ref{fig:gmapFT_3T}(d)-(f). In particular, qualitatively different behaviors can be seen between the OP case and the OD ones. At the optimum doping, as shown in Fig.~\ref{fig:gmapFT_3T}(d), a clear energy-dependent dispersion at $\mathbf Q$ and $2\mathbf Q$ can be seen for bias potential $0.2 \gtrsim \omega \gtrsim 0.15$, indicating the presence of a CDW order. Interesting, this is also the energy range that a dominant $s$+$s'$ form factor is obtained as discussed above. For bias smaller than $0.15$, the $2\mathbf Q$ peaks are reduced, and there seems to be some energy dispersion. This bipartite structure become more apparent for the OD samples.  In Figs.~\ref{fig:gmapFT_3T}(e) and (f) the feature at the fundamental wave vector $\mathbf{Q}a/2\pi =0.125$ merges with with the low $k$ feature and no longer identifiable. The intensity at $2\mathbf{Q}$ also becomes weaker weaker, and an energy dispersion seems to be developing for $|\omega|<0.15$. Although the CB-h vortex of both OP and OD cases exhibit a CDW order, as can be directly seen in the real-space configurations (Figs.~S12(b) and (c) in the supplemental material), the apparently stronger dispersion at low energy in the OD case could result from a more prominent QPI effect. 

Our results also shed light on the seemingly conflicting interpretations of the experiment in Ref.~\cite{Renner21}. While their STM imaging on a sample of dopant density of $\sim 0.2$ in a 3T magnetic field found a checkerboard pattern inside the vortex halo, an energy dispersion similar to Fig.~\ref{fig:gmapFT_3T}(e) was also observed, which prompted the authors to suggest the QPI origin, instead of CDW, for the observed checkerboard pattern. However, our calculations of the CB-h vortex state clearly show that a dispersive feature of the conductance map could also result from an intertwined PDW/CDW state.
This bipartite structure is  consistent with the previous work \cite{Machida16} that the PDW is related to the antinode or the pseudogap with higher energy than the BCS pairing near the nodal region which dominates in QPI.


\begin{figure}
	\includegraphics[width=1.0\columnwidth]{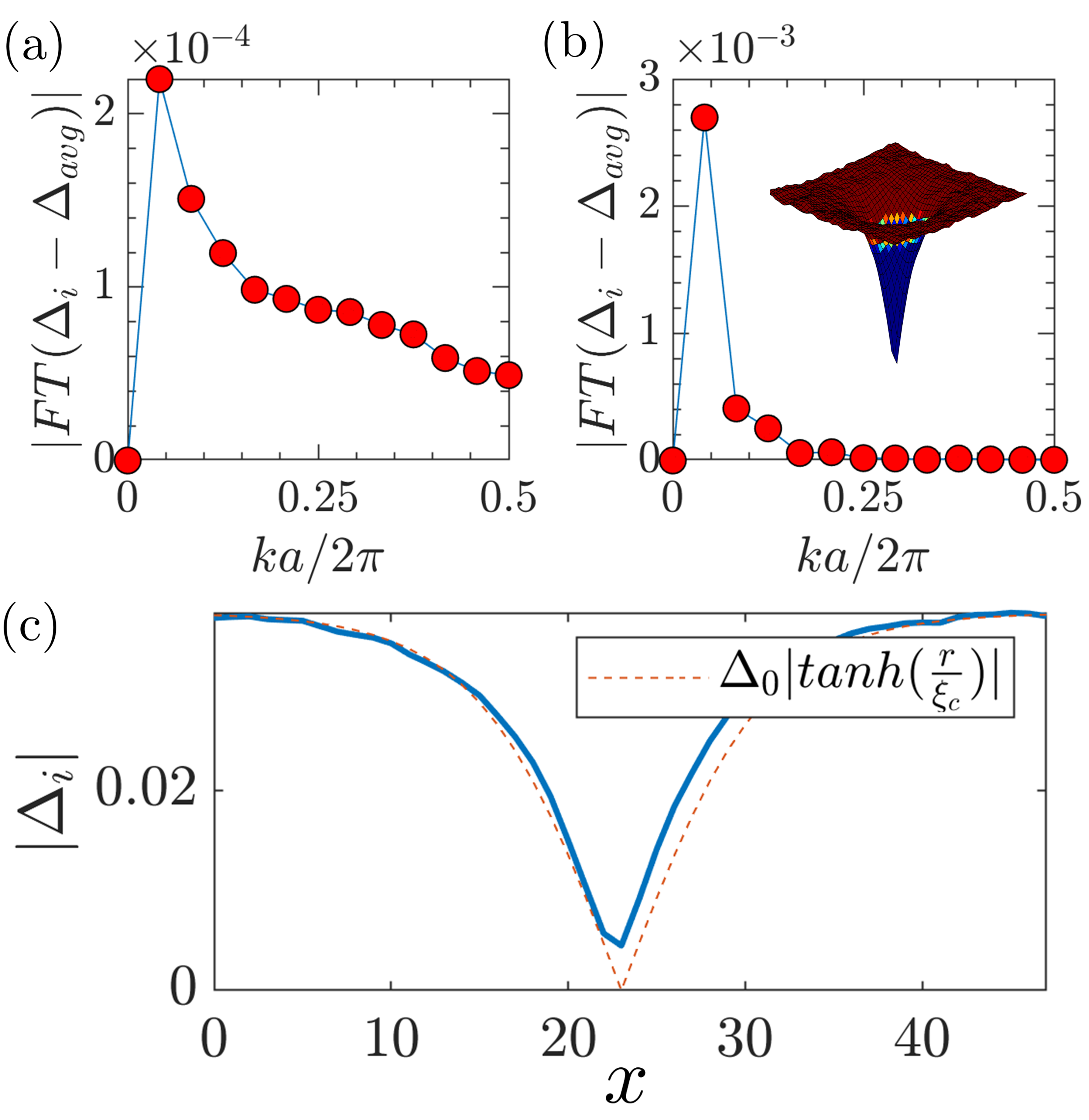}
	\caption{ The discrete Fourier transform  of the RMFT and non-correlation mean field for pure $d$-wave vortex state.
		(a) The DFT of plain-vortex solution shown in Fig.~\ref{fig:0}(a).
		(b) The DFT of traditional $d$-wave vortex without including the strong correlation. The inset shows the 3D map of pairing amplitude $|\Delta_i|$.
		(c) Positional dependence of the pairing amplitude $|\Delta_i|$ of the traditional $d$-wave vortex state for B = 5.96T and $\delta = 0.16$. A line cut along $x$-axis passing through the center of the vortex at core (23, 23). The fitting curve here used $\xi_c \sim 8a$.
	} 		
	\label{fig:figend}  \
\end{figure}

\subsection{ CB-halo State Induced by Vortex }

There are many experimental evidences~\cite{Li07,He11,Hashimoto14,Hamidian16,Norman18,Rajasekaran18,Edkins19} and theoretical works~\cite{Himeda02,Raczkowski07,Aligia09,Yang09,Loder11,PALee14,Corboz14,Tu16,Choubey17,Tu19,Choubey20,Choubey21} that support the presence of PDW states in the UD and OP cuprates without a  magnetic field. Thus it is easier to foresee that inside a vortex where both hole density and pairing amplitude are reduced so that the situation is more like an UD regime that PDW states will become energetically favorable. By comparison the plain-vortex solution with the CB-h state we found another reason for the strong coupling of PDW state with a vortex. 

As mentioned earlier the order parameter of a PDW state is a Cooper pair with a finite total momentum $\mathbf{Q}$, $\langle c_{\mathbf{k}+\frac{\mathbf{Q}}{2}} c_{-\mathbf{k}+\frac{\mathbf{Q}}{2}} \rangle$ , or the FFLO order.

Thus the pairing amplitude is non-uniform. Since inside the vortex, the pairing amplitude decays exponentially from the center of the vortex as shown in Fig. 1(b). Its FT, shown in Fig. 7(a), decays rapidly as $k$ increases but with a small peak  around $ka/2\pi=0.3$. Its inverse is close to the radius of the vortex about  $3a$. Thus all these momenta will provide a FFLO order.  The corresponding curve for CB-h state shown in Fig. 3(f), has a similar shape except a peak at  $ka/2\pi=0.125$. But notice the magnitude is of order $10^{-4}$, thus two cases are quite close.  Further more the values of $|\sum_k\langle c_{\mathbf{k}+\frac{\mathbf{Q}}{2}} c_{-\mathbf{k}+\frac{\mathbf{Q}}{2}} \rangle|$ for $\mathbf{Q}a/2\pi=(0125,0)$ and $(0.25,0)$ are the same, 0.016 and 0.0096 for plain-vortex and CB-h states respectively.  Details are given in the SM. This surprising result has two implications. The plain-vortex already has FFLO orders with a range of momentum, possibly smaller than $ka/2\pi=0.3$.  Thus the choice of one particular order such as $ka/2\pi=0.125$ only causes a very minor effect on the vortex. This may explain why the CB-h state has almost same energy as plain-vortex state (Table SI in SM) and many properties are similar. On the other hand the period of the PDW state or CB is not determined a priori. Just like the calculations of the ground state without magnetic field for the Hubbard~\cite{Zheng17} or $t$-$J$ model~\cite{Tu16}, PDW states with different periods have almost the same energy. 

To support our argument above, we also calculate a traditional $d$-wave vortex without including the strong correlation and applying RMFT.  We choose a simple $d$-wave pairing model with the pairing amplitude same as $\delta=0.16$ of our plain-vortex solution shown in Fig.~\ref{fig:0}(a).  
The discrete Fourier transform of pairing amplitude of such traditional vortex at 5.98T are shown in Figs. \ref{fig:figend}(b). 
The inset of Fig.\ref{fig:figend}(b) shows the 3D map of pairing amplitude.
Due to the large radius of the vortex shown in Fig.\ref{fig:figend}(c) where the core size in the fitting curve $\xi_c \sim 8a$ is used.
The  Fourier transform only shows a very large peak at small k and decays smoothly with k. This is very different from Fig. \ref{fig:figend}(a) with a small peak around $ka/2\pi \sim 0.3$.  It turns out the values FFLO order $|\sum_k\langle c_{\mathbf{k}+\frac{\mathbf{Q}}{2}} c_{-\mathbf{k}+\frac{\mathbf{Q}}{2}} \rangle|$ at $Qa/2\pi=0.125$ is about two times smaller than the plain vortex  and CB-h solutions. 
The smaller size of the vortex core due to strong correlation has a much larger coupling with FFLO order at larger $Qa/2\pi=0.125$.


\section{Summary}

\label{sec:conclusion}

By taking into account the renormalization effect due to the strong correlation in the $t$-$t’$-$J$ model, the vortex structure in a d-wave SC state is calculated for several hole densities with magnetic field in the range of 3T to 12T. Two states with almost the same energy are used to understand puzzles found by experiments. Tunneling conductance is analyzed for the plain-vortex and CB-h states to compare with the STM experiments~\cite{Machida16,Edkins19,Renner21}. The absence of ZBCP in vortex core for UD and OP cuprates~\cite{Machida16,Edkins19} but not for heavily OD samples is easily explained. The presence of PDW states in CB-h solution will suppress the conductance peak with a subgap structure. However, even if there were no PDW presence, the ZBCP is greatly  reduced in conductance spectra due to the influence of $d_{x^2-y^2}$ orbital symmetry of the copper on the STM tip. Only for OD samples, the greatly reduced ZBCP is still visible as shown in the experiment of Ref.~\cite{Renner21}. The CB-h state in the vortex halo has a bidirectional PDW state with modulation period of 8a and it shows clear $s$+$s'$ form factors at wave vectors $Q_x=(1/8,0) 2\pi/a$ and $2Q_x$ as well as peaks in y direction as reported by Ref.~\cite{Edkins19}. 

The presence of both QPI at low energy and CDW at higher energy of the bipartite structure reported first by Ref~\cite{Machida16} is also a distinct property of the CB-h state. In these states the interplay between QPI and CDW depend on the hole density. For UD and OP cases, the non-dispersive CDW shows strong presence in the conductance map, $g(\mathbf k, \omega)$, for energy around or greater than half of the SC gap but it has a QPl-like energy dispersion at lower energy. 
The CDW signal becomes weaker as hole density increases toward the OD regime. 
The QPI effect is mostly contributed by the Bogoliubov quasiparticles near the nodes of SC gap; hence they are at lower energy. 
By contrast, the  CDW is related to the PDW state at antinodes with higher energy close to pseudogap. Such a bipartite structure in the solutions without magnetic field \cite{Tu19} is nicely reflected in the vortex structure.

Just as reported in Ref.~\cite{Edkins19} our CB-h state has stronger $s$+$s'$ form factors than $d$ form factors. However our state has the presence of SDW order intertwined with PDW and CDW orders while the STM experiment did not provide the magnetic information. If there were no SDW order, the $d$ form factors will dominate (This part is not included in this paper). Although there are several experiments indicating the magnetic signal inside the vortex Ref.~\cite{Lake01,Lake02}, more direct evidence for the presence of SDW is welcome. 

Our result provided a new insight about the coupling between the bidirectional PDW states or CB  with a vortex. The pairing order parameter inside a vortex is a function of the position, its DFT are FFLO orders $\langle c_{\mathbf k + \mathbf Q/2, \uparrow} c_{-\mathbf k + \mathbf Q/2, \downarrow} \rangle$. But for usual superconductors, the long coherence length will have a vortex with large radius, thus only small-$\mathbf Q$ FFLO orders are important. However, for cuprates, the strong correlation makes the vortex very small with a radius of three unit cells only. Thus large-$\mathbf Q$ FFLO orders are also important. Thus vortex could then easily induce the specific-$\mathbf Q$ PDW states already preferred in the cuprates without magnetic field. 

In our previous works~\cite{Tu16}, we have shown that unidirectional PDW states with different periods all have similar energies in this approach using RMFT for the $t$-$J$ model.  This is also found by more accurate numerical methods \cite{Zheng17} for the Hubbard model. Thus it is difficult to identify what value of the period is more preferred for the PDW ground state from numerical calculations. Or the preference of particular period like $8a$ is just not included in the model.  A recent experiment has indicated that the electron-phonon coupling may favor CDW to have a period $4a$ than $6a$ \cite{Huang21}. A consequence of the presence of FFLO orders in the vortex is that the heavily doped samples, which have no PDW states reported so far and also not favored in the theoretical calculations\cite{Tu16, Choubey17}, will be induced by the magnetic field.

\appendix

\section{Renormalized mean-field theory}

\label{sec:RMFT}

Here we present details of the RMFT calculations for the $t$-$t'$-$J$ model. In Gutzwiller's original calculation, the strong on-site electron correlation effectively leads to a reduced inter-site electron hopping~\cite{Gutzwiller1963}. By generalizing this approach to also account for the inter-site exchange interaction~\cite{Himeda1999,Ogata03,springerthesis}, the $t$-$J$ Hamiltonian in Eq.~(\ref{eq:H_tJ}) becomes
\begin{eqnarray}
	& & \hat H_{\rm renorm}=-\sum_{ij,\sigma} g^t_{ij\sigma} \, t_{ij} \left(\hat{c}^\dagger_{i\sigma} \hat{c}_{j\sigma} +  \mbox{h.c.} \right) \\
	& & \qquad +\sum_{\langle i,j\rangle}J\Bigg [ g^{s,z}_{ij} \hat{S}^{z}_i \hat{S}^{z}_j+g^{s,xy}_{ij}\Bigg(\frac{\hat{S}^{+}_i \hat{S}^{-}_j + \hat{S}^{-}_i \hat{S}^{+}_j}{2}\Bigg)\Bigg], \nonumber
	\label{eq:H_renorm}
\end{eqnarray}
where the various $g$ factors, also known as the Gutzwiller factors, depend on the local order parameters Eq.~(\ref{eq:delta})-(\ref{eq:pairing}). Specifically, the electron hopping is renormalized by the factor $g^t_{ij\sigma}=g^t_{i\sigma}g^t_{j\sigma}$, where the on-site Gutzwiller factor is 
\begin{eqnarray}
	g^t_{i\sigma}=\sqrt{ \frac{2\delta^v_i( 1 - \delta^v_i)}{1-(\delta_i^v)^2 + 4 (m_i^v)^2} 
	\frac{1 + \delta^v_i + \sigma 2 m_i^v }{1 + \delta^v_i - \sigma 2 m^v_i }   }.
\end{eqnarray}
In this work, we focus on collinear SDW and assume that its magnetization points along the $\pm z$ direction, the SU(2) rotational symmetry is explicitly broken. The renormalization of the transverse $xy$ exchange is given by $g^{s,xy}_{ij}=g^{s,xy}_i g^{s,xy}_j$, where
\begin{eqnarray}
	g^{s,xy}_i=\frac{2( 1 - \delta^v_i)}{1-(\delta^v_i)^2 + 4 (m^v_i)^2}.
\end{eqnarray}
The renormalization factor for the exchange interaction along the $z$ direction is related to that of the $xy$ direction via
\begin{eqnarray}
	g^{s,z}_{ij}=g^{s,xy}_{ij} \frac{2 \left[  ({\bar{\Delta}^v_{ij})^{2}}-(\bar{\chi}^v_{ij})^{2}  \right]
	- 4m^v_i m^v_j X_{ij}^2 } {2 \left[  (\bar{\Delta}^v_{ij})^{2}-(\bar{\chi}^v_{ij})^{2} \right] - 4m^v_i m^v_j  },
\end{eqnarray}
where $\bar{\Delta}^v_{ij} = \sum_\sigma {\Delta^v_{ij\sigma}}/2$ and  $\bar{\chi}^v_{ij} = \sum_\sigma {\chi^v_{ij\sigma}}/{2}$, and the $X$-factor is defined as
\begin{eqnarray}
	X_{ij} = 1 + \frac{12(1 - \delta^v_i)(1-\delta^v_j)}{\sqrt{ \left[ 1-(\delta^v_i)^2+4 (m^v_i)^2 \right] \left[ 1- (\delta^v_j)^2+4 (m^v_j)^2 \right] }} \nonumber \\
\end{eqnarray}
Note that in the absence of magnetic order $m_i = 0$, the exchange renormalization factors reduce to $g_{ij}^{s,z}=g_{ij}^{s,xy}$, indicating an intact SU(2) rotation symmetry.

The effective mean-field Hamiltonian can be obtained by minimizing the total energy of the system computed from a Slater-determinant state $|\Phi_0\rangle$, which is to be self-consistently determined. As the minimization is subject to the constraints of fixed electron number $\sum_i n_i = N_e$ and of a normalized variational wave function $\langle \Psi_0 | \Psi_0 \rangle = 1$, we introduce two Lagrangian multipliers and define the following effective energy
\begin{eqnarray}
	& & W = \langle \Psi_0| \hat{H}_{\rm renorm} |\Psi_0\rangle - \lambda( \langle \Psi_0| \Psi_0 \rangle  - 1) \nonumber \\
	& & \qquad - \mu \Bigl(\sum_i n_i - N_e \Bigr),
\end{eqnarray}
The optimization with respect to the unprojected wave function $\partial W / \partial |\Psi_0 \rangle$ leads to the following effective TB-BdG Hamiltonian
\begin{eqnarray}
	& & \hat H_{\rm eff} = \sum_{ij,\sigma} \Bigl(\frac{\partial W}{\partial  \chi^v_{ij\sigma}} \hat{c}^\dagger_{i\sigma} \hat{c}_{j\sigma} + \mbox{h.c.} \Bigr) \\
	& & \quad + \sum_{\langle ij\rangle ,\sigma} \Bigl(\frac{\partial W}{\partial  \Delta^v_{ij\sigma}} \hat{c}_{i\sigma} \hat{c}_{j\bar \sigma} 
	+ \mbox{h.c.} \Bigr)  +\sum_{i\sigma} \frac{\partial W}{\partial  n_{i\sigma}} \hat{n}_{i\sigma}, \nonumber
	\label{eq:Hamiltonian3}
\end{eqnarray}
which is equivalent to Eq.~(\ref{eq:H_eff}) in the main text. The above expression also provides the definition for the effective hopping, pairing, and on-site energy parameters, given in Eq.~(\ref{eq:model_param}). Explicitly, they are given by
\begin{eqnarray}
	\label{eq:Tij}
	\mathcal{T}_{ij\sigma}= -g^t_{ij\sigma}t_{ij}  -J \Bigl( \frac{g^{s,z}_{ij}}{4}
	+ \frac{g^{s,xy}_{ij}}{2} \frac{\chi^{v*}_{ij\bar \sigma }}{\chi^{v*}_{ij,\sigma}} \Bigr) \chi_{ij\sigma}^{v*}		
 + \left[ \frac{\partial W}{\partial \chi^v_{ij\sigma}} \right]_g, \nonumber  \\ 
\end{eqnarray}
\begin{eqnarray}
	\label{eq:Dij}
	\mathcal{D}_{ij\sigma}= -J \Bigl( \frac{g^{s,z}_{ij}}{4} 
	+\frac{g^{s,xy}_{ij}}{2} \frac{\Delta^{v*}_{ij\bar \sigma }}{\Delta^{v*}_{ij,\sigma}} \Bigr) \Delta_{ij\sigma}^{v*} 
	+ \left[ \frac{\partial W}{\partial \Delta^v_{ij\sigma}} \right]_g, \nonumber \\
\end{eqnarray}
\begin{eqnarray}
	\label{eq:mu_i}
	\mu_{i\sigma}=-\mu + \frac{1}{2}\sigma \sum_j g^{s,z}_{ij}Jm^v_j
	+ \left[ \frac{\partial W}{\partial n_{i\sigma}} \right]_g.
\end{eqnarray}
Here the terms $\left[\frac{\partial W}{\partial O}\right]_g$ refer to the derivative of $W$ with respect to the order parameter $O$ via all possible Gutzwiller factor $g$,
\begin{eqnarray}
	 \left[\frac{\partial W}{\partial \Delta^v_{ij\sigma}}\right]_g 
	= \frac{\partial W}{\partial g^{s,z}_{ij}}\frac{\partial g^{s,z}_{ij}}{\partial \Delta^v_{ij\sigma}}, \qquad
	\left[\frac{\partial W}{\partial \chi^v_{ij\sigma}}\right]_g 
	= \frac{\partial W}{\partial g^{s,z}_{ij}}\frac{\partial g^{s,z}_{ij}}{\partial \chi^v_{ij\sigma}}, \quad  \nonumber 
\end{eqnarray}
\begin{eqnarray}
	\left[\frac{\partial W}{\partial n_{i\sigma}}\right]_g = \frac{\partial W}{\partial g^{s,xy}_{ij}}\frac{\partial g^{s,xy}_{ij}}{\partial n_{i\sigma}} 
	+ \frac{\partial W}{\partial g^{s,z}_{ij}}\frac{\partial g^{s,z}_{ij}}{\partial n_{i\sigma}}
	+ \sum_{\sigma^{\prime}}\frac{\partial W}{\partial g^{t}_{ij\sigma^{\prime}}}\frac{\partial g^{t}_{ij\sigma^{\prime}}}{\partial n_{i\sigma}}. 
	\nonumber 
\end{eqnarray}


Computing the various local order-parameters in Eq.~(\ref{eq:delta})--(\ref{eq:pairing}) from the ground state of the effective Hamiltonian $\hat{H}_{\rm eff}$ is a central step in the RMFT iteration. In most calculations for a system of $10^5$ lattice sites, up to $10^3$ iterations are routinely required before convergence can be reached. Consequently, a highly-efficient GPU-based kernel polynomial method (KPM), discussed in Appendix~\ref{sec:kpm}, is used for the optimization of the various local orders $\delta_{i}, m_i, \chi_{ij\sigma}$, and $\Delta_{ij\sigma}$; the total number of these parameters is of the order of $10^5$.

\section{Kernel polynomial method}

\label{sec:kpm}

The kernel polynomial method (KPM)~\cite{Weisse06} and the technique of automatic differentiation~\cite{Barros13,ZWang18} play a crucial role in our large-scale RMFT calculations. The basic formulation of these techniques is outlined in this section.  Conventional KPM provides an efficient approach to computing the correlation function, or single-particle density matrix, $\rho_{i\alpha, j\beta} = \langle c^\dagger_{j\beta} c^{\,}_{i \alpha} \rangle$ of large sparse tight-binding matrices. For superconducting systems, one needs to include the pairing term, as well as calculate the anomalous correlation $\langle c^{\,}_{i\sigma} c^{\,}_{j \bar{\sigma}} \rangle$. A straightforward generalization of KPM by expressing the Hamiltonian in terms of Nambu spinors can be used to compute these additional correlation functions. To this end, we define the two-component Nambu spinor as
\begin{eqnarray}
	\hat{\mathbf d}_i = \left(\begin{array}{c} \hat{d}^{\,}_{i, 1} \\ \hat{d}^{\,}_{i, 2} \end{array} \right) = \left(\begin{array}{c} \hat{c}^{\,}_{i \uparrow} \\ \hat{c}^\dagger_{i \downarrow} \end{array}\right)
\end{eqnarray}
The effective Hamiltonian~(\ref{eq:H_eff}) can then be expressed as
\begin{eqnarray}
	& & \hat{H}_{\rm eff} = \sum_{ij} \hat{\mathbf d}^\dagger_i
	\left(\begin{array}{cc}
		\mathcal{T}_{ij,\uparrow}-\mu_{i\uparrow}\delta_{ij} & \mathcal{D}_{ij\uparrow} \\
		\mathcal{D}^{*}_{ji\downarrow}	  & -\mathcal{T}^*_{ij,\downarrow}+\mu_{i\downarrow}\delta_{ij}
	\end{array} \right) \hat{\mathbf d}^{\,}_j, \nonumber \\
	& & \qquad = \sum_{i\alpha, j\beta}  h^{\,}_{i\alpha, j\beta} \hat{d}^\dagger_{i\alpha} \hat{d}^{\,}_{j\beta}
	= \sum_{IJ} h^{\,}_{IJ} \,\hat{d}^{\dagger}_I \, \hat{d}^{\,}_J.
\end{eqnarray}
Here we have introduced the single-particle Hamiltonian matrix $h_{i\alpha, j\beta}$, where $\alpha, \beta = 1, 2$ is the Nambu index. For convenience, we have further introduced the short-hand notation $I = (i, \alpha)$, $J = (j, \beta)$, $\cdots$, and so on. The correlation function, or single-particle density matrix, of the $d$-fermions is defined as
\begin{eqnarray}
	\rho_{IJ} \equiv \langle \hat{d}^\dagger_I \hat{d}^{\,}_J \rangle
\end{eqnarray}
It is worth noting that $\rho_{IJ}$ now includes both the normal correlation function $\langle \hat{c}^\dagger_{i\sigma} \hat{c}^{\,}_{j \sigma} \rangle$ and the anomalous $\langle \hat{c}^{\,}_{i\uparrow} \hat{c}^{\,}_{j \downarrow} \rangle$. Here $\langle \hat{O} \rangle = {\rm Tr}(e^{-\beta\hat{H}_{\rm eff}} \hat{O}) / Z$ denotes the expectation value computed from the effective TB-BdG Hamiltonian, where $Z = {\rm Tr} e^{-\beta \hat{H}_{\rm eff}}$ is the partition function of the quadratic effective Hamiltonian. To compute the correlation functions, we consider the energy of the ground state:
\begin{eqnarray}
	\mathcal{E} = \langle   \hat{H}_{\rm eff}  \rangle = \sum_{IJ} h_{IJ} \langle \hat{d}^{\dagger}_I \, \hat{d}^{\,}_J \rangle.
\end{eqnarray} 
It is then easy to see that the density matrix is then given by the derivative
\begin{eqnarray}
	\rho_{IJ} = \frac{\partial \mathcal{E}}{\partial h_{IJ}}.
\end{eqnarray}
Next we introduce the total density of states (DOS) $\rho(\epsilon)$ of the system and express the energy as
\begin{eqnarray}
	\mathcal{E} = \int \rho(\epsilon) \, f(\epsilon) d\epsilon, 
\end{eqnarray}
where $f(\epsilon) = -T \log [1 + e^{-(\epsilon-\mu)/T}]$. The central step of KPM is to approximate the DOS as a Chebyshev polynomial series,
\begin{eqnarray}
	\label{eq:rho_KPM}
	\rho(\epsilon) = \frac{1}{\pi \sqrt{1 - \epsilon^2}} \sum_{m=0}^{M-1} (2 - \delta_{0, m})  \mu_m\, T_m(\epsilon),
\end{eqnarray}
where $T_m(x)$ are Chebyshev polynomials, and $\mu_m$ are the expansion coefficients. The expansion is valid only when all eigenvalues of $H_{IJ}$ have magnitude less than one. This can in general be achieved through a simple shifting and rescaling of the Hamiltonian. Moreover, damping coefficients $g_m$ are often introduced to reduce the unwanted artificial Gibbs oscillations. Substituting $\rho(\epsilon)$ into the free energy expression gives
\begin{eqnarray}
	\label{eq:F_KPM}
	\mathcal{E} = \sum_{m=0}^{M-1} C_m\, \mu_m,
\end{eqnarray}
where coefficients $C_m = (2 - \delta_{0,m}) g_m \int_{-1}^{1} \frac{T_m(\epsilon) f(\epsilon)}{\pi \sqrt{1-\epsilon^2}} d\epsilon$ are independent of the Hamiltonian and may be efficiently evaluated using Chebyshev-Gauss quadrature. 

In KPM, the calculation of the Chebyshev moments $\mu_m = {\rm Tr}  \, T_m(h)$ is approximated by an ensemble average $\mu_m = \langle T_m(h) \rangle =  \frac{1}{R}\sum_{\ell=1}^{R} r_\ell^\dagger \cdot h \cdot r_\ell$ over random normalized column vectors $r$~\cite{Silver94}. Taking advantage of the recursive relation of Chebyshev polynomials: $T_m(h) = 2 H \cdot T_{m-1}(h) - T_{m-2}(h)$, the moments can be evaluated recursively as follows:
\begin{eqnarray}
	\mu_m = r^\dagger \cdot \alpha_m,
\end{eqnarray}
where $r$ is a random vectors with complex elements drawn from the uniform distribution $|r_I|^2 = 1$. The random vectors $\alpha_m$ are given by
\begin{eqnarray}
	\label{eq:alpha}
	\alpha_m = \left\{\begin{array}{lcr} 
	r,  & \qquad & m = 0 \\
	h \cdot r, & \qquad & m=1 \\
	2 h \cdot \alpha_{m-1} - \alpha_{m-2}, & \qquad & m>1 \\
	\end{array}\right.
\end{eqnarray}
The above recursion relation also indicates that evaluation of $\mu_m$ that are required for computing $\mathcal{E}$ only involves matrix-vector products. For sparse matrix $h$ with $\mathcal{O}(N)$ elements, this requires only $\mathcal{O}(M N)$ operations, where $M$ is the number of Chebyshev polynomials. On the other hand, even with the efficient algorithm for $\mathcal{E}$, a naive calculation of the derivatives $\partial \mathcal{E}/\partial h_{IJ}$ based on finite difference approximation is not only inefficient but also inaccurate. The computational cost of finite difference is similar to the KPM-based Monte Carlo method with local updates.

To circumvent this difficulty, we employ the technique of automatic differentiation with reverse accumulation. Instead of directly using Eq.~(\ref{eq:F_KPM}), the trick is to view $\mathcal{E}$ as a function of vectors $\alpha_m$ and write
\begin{eqnarray}
	\label{eq:dFdH}
	\frac{\partial \mathcal{E}}{\partial h_{IJ}} = \sum_{m=0}^{M-1} \frac{\partial \mathcal{E}}{\partial \alpha_{m, K}} \frac{\partial \alpha_{m, K}}{\partial h_{IJ}},
\end{eqnarray}
Here $\alpha_{m, K}$ denotes the $K$-th component of vector $\alpha_m$, and summation over the repeated index $K$ is assumed. Using Eq.~(\ref{eq:alpha}), we have 
\begin{eqnarray}
	\label{eq:dalphadH}
	& & \frac{\partial \alpha_{0, K}}{\partial h_{IJ}} = 0, \qquad \frac{\partial \alpha_{1, K}}{\partial h_{IJ}} = \delta_{IK}\, \alpha_{0, J}, \nonumber \\
	& & \frac{\partial \alpha_{m, K}}{\partial h_{IJ} } = 2 \delta_{IK}\, \alpha_{m-1, J}  \quad (m>1)
\end{eqnarray}
The expression of $\partial \mathcal{E}/\partial h_{IJ}$ can be simplified by introducing a new set of random vectors:
\begin{eqnarray}
	\beta_m \equiv \frac{\partial \mathcal{E}}{\partial \alpha_{m+1}},
\end{eqnarray}
From Eqs.~(\ref{eq:dFdH}) and~(\ref{eq:dalphadH}) , we obtain
\begin{eqnarray}
	\frac{\partial \mathcal{E}}{\partial h_{IJ}} = \beta_{0, I} \,\alpha_{0, J} + 2 \sum_{m=1}^{M-2} \beta_{m, I} \, \alpha_{m, J}.
\end{eqnarray}
Remarkably, the vectors $\beta_m$ can also be computed recursively. To this end, we note that the recursion relation~(\ref{eq:alpha}) implies that $\mathcal{E}$ depends on $\alpha_m$ through three paths:
\begin{eqnarray}
	& & \frac{\partial \mathcal{E}}{\partial \alpha_{m, K}} = \frac{\partial\mathcal{E}}{\partial \mu_m} \frac{\partial \mu_m}{\partial \alpha_{m, K}}  \\
	& & \qquad + \frac{\partial \mathcal{E}}{\partial \alpha_{m+1, L}} \frac{\partial \alpha_{m+1, L}}{\partial \alpha_{m, K}}
	+ \frac{\partial \mathcal{E}}{\partial \alpha_{m+2, L}} \frac{\partial \alpha_{m+2, L}}{\partial \alpha_{m, K}}. \nonumber
\end{eqnarray}
The various terms above can be straightforwardly calculated:
\begin{eqnarray}
	& & \frac{\partial \mathcal{E}}{\partial \mu_m} = C_m, \quad \frac{\partial \mu_m}{\partial \alpha_{m, K}} = r^*_K, \nonumber \\
	& & \frac{\partial \alpha_{m+1, L}}{\partial \alpha_{m, K}} = 2 h_{LK}, \quad 
	\frac{\partial \alpha_{m+2, L}}{\partial \alpha_{m, K}} = -\delta_{LK}.
\end{eqnarray}
Consequently, 
\begin{eqnarray}
	\beta_m = C_{m+1} \, r^\dagger + 2 \beta_{m+1} \cdot h - \beta_{m+2}, 
\end{eqnarray}
for $m < M-1$.
As in standard KPM, there are two independent sources of errors in our method~\cite{Weisse06,Barros13}: the truncation of the Chebyshev series at order $M-1$, and the stochastic estimation of the moments using finite number $R$ of random vectors.  The performance of the stochastic estimation can be further improved using correlated random vectors based on the probing method~\cite{Tang12}. 
Most simulations discussed in the main text were done on a $120\times 120$ triangular lattice. The number of Chebyshev polynomials used in the simulations is in the range of $M = 1000$ to $2000$. The number of correlated random vectors used is $R = 64$ to $144$.

\section{Calculation of local density of states and magnetic super-cell method}

\label{sec:ldos}

Once the various local order parameters $\delta_{i},  m_i, \chi_{ij\sigma}$, and $\Delta_{ij\sigma}$ are obtained in the KPM-RMFT calculations discussed above, we combine the exact diagonalization (ED) with the supercell method to compute the electron Green's function and the local density of states. It is worth noting that although the ED of large TB-BdG matrices is rather time-consuming due to its $\mathcal{O}(N^3)$ scaling, here only one diagonalization is required for each set of the local order parameters. Moreover, the eigen wave functions obtained from the ED allow us to utilize the super-cell method to further increase the resolution of the Green's functions and density of states. To this end, we first express the effective mean-field Hamiltonian in the standard Bogoliubov-de Gennes (BdG) form:
\begin{eqnarray}
\label{eq:diagbdg}
	 & &H_{\rm eff}
	 \begin{pmatrix}
		u^n_j \\
		v^n_j  
	\end{pmatrix} \nonumber \\
	& & \quad =\sum_j
	\begin{pmatrix}
		\mathcal{T}_{ij,\uparrow}-\mu_{i\uparrow}\delta_{ij} & \mathcal{D}_{ij\uparrow} \\
		\mathcal{D}^{*}_{ji\downarrow}	  & -\mathcal{T}^*_{ij,\downarrow}+\mu_{i\downarrow}\delta_{ij}
	\end{pmatrix}
	\begin{pmatrix}
		u^n_j \\
		v^n_j  
	\end{pmatrix} \nonumber \\
	& & \quad = E_n
	\begin{pmatrix}
		u^n_i \\
		v^n_i 
	\end{pmatrix},
\end{eqnarray}
where the effective hopping, pairing, and on-site energy parameters are defined in Eq.~(\ref{eq:Tij}), (\ref{eq:Dij}), and (\ref{eq:mu_i}), respectively, and $(u^n_i, v^n_i)$  and $E_n$ are the eigenvectors and eigenvalues. 

Next the effect of a magnetic field on a tight-binding model is described using the standard Peierls substitution, which adds a phase factor to the bare hopping constant: $t_{ij} \to t_{ij} \, e^{i\phi_{ij}}$, where $\phi_{ij} = \frac{-\pi}{\Phi_0} \int_{\mathbf{r}_i}^{\mathbf{r}_j} \mathbf {A}(\mathbf{r}_i)\cdot d\mathbf{r}_i $, $\Phi_0=\frac{hc}{2e}$ is the superconducting flux quantum, and $\mathbf A(\mathbf r)$ is the vector potential that generates the magnetic field, i.e. $\mathbf {B}=\nabla \times \mathbf {A}$. A Landau gauge $\mathbf {A} = B (0,x)$ is used for magnetic fields along the $z$ direction. The field strength is set to be $B=2\Phi_0/N_xN_y$, where $N_x\times N_y$ is the size of the system.

Given the TB-BdG equation~(\ref{eq:diagbdg}), the magnetic super-cell method~\cite{Schmid10} allows us to extend the effective linear dimension of the system to $M_x \times N_x$ and $M_y \times N_y$, along the $x$ and $y$ directions, respectively. Here $M_x$ and $M_y$ are the number of super-cells in the two orthogonal directions. From the magnetic Bloch theorem~\cite{Kohn1959}, the boundary conditions for the eigenstates now becomes
\begin{equation}\label{eq:matrixeqn}
	\begin{pmatrix}
		u^n_{\mathbf{k}}(\mathscr{T}_{\mathbf{R}} \, \tilde{\mathbf{r}} _i) &  \\
		v^n_{\mathbf{k}}(\mathscr{T}_{\mathbf{R}} \, \tilde{\mathbf{r}} _i) & 
	\end{pmatrix}
	=e^{-i\mathbf{k}\cdot \mathbf{R}_i }
	\begin{pmatrix}
		e^{-i\xi({\mathbf r}_i, \mathbf{ R} )/2}
		u_{\mathbf{k}}( \tilde{\mathbf{r}} _i) &  \\
		e^{i\xi({\mathbf r}_i, \mathbf{ R} )/2}
		v_{\mathbf{k}}( \tilde{\mathbf{r}} _i) & 
	\end{pmatrix},
\end{equation}
The magnetic super-cell of the enlarged lattice is indexed by $\boldsymbol{R} = \mathbf{R}_{n_x, n_y} =n_x N_x \mathbf{\hat{x}} + n_y N_y \mathbf{\hat{y}}$, the vector $\tilde{\mathbf{r}}_i$ denotes the relative position vector within a given magnetic unit cell, $\mathbf{k} = \mathbf{k}_{m_x, m_y} = \frac{2\pi m_x}{M_x N_x} \mathbf{\hat{x}} + \frac{2\pi m_y}{M_y N_y} \mathbf{\hat{y}}$ is the momentum of the enlarged lattice, and the phase $\xi(\mathbf{r}_i,\mathbf{R}) = \frac{2\pi}{\Phi_0}\mathbf{A}(\mathbf{R})\cdot \mathbf{r}_i -4\pi n_x n_y $. 
We have also defined the magnetic translation operator $\mathscr{T}_{\mathbf{R}} = \exp[-i\mathbf{R}\cdot (\mathbf{k}+q{\mathbf A}/c\hbar ) ]$~\cite{Brown1964} which commutes with the BdG Hamiltonian.

From the eigen-functions $\{ u^n_i, v^n_i \}$ of the TB-BdG Hamiltonian, the lattice Green's function in frequency-domain is given by \cite{Gijcomplex}
\begin{eqnarray}\label{eq:Gij}
	G_{ij}(\omega) =  \sum_n \left(  \frac{ g^t_{ij,\uparrow}{u_i^n u_j^{n*}}}{\omega - E_n + i 0^+} 
	+ \frac{ g^t_{ij,\downarrow}{v_i^{n*} v_j^{n}}}{\omega + E_n + i 0^+}  \right). \nonumber \\
\end{eqnarray}
The local density of states and other experimental measurements can be obtained from the lattice Green's functions.

\begin{acknowledgments}
YH Liu and TK Lee are grateful for the support of the Ministry of Science and Technology, Taiwan under Grants No. 110-2112-M-110-018. GW Chern is partially supported by the Center for Materials Theory as a part of the Computational Materials Science (CMS) program, funded by the US Department of Energy, Office of Science, Basic Energy Sciences, Materials Sciences and Engineering Division. The authors also acknowledge the computer support of the National Center for High-Performance Computing in Taiwan and the Academia Sinica Grid-computing Center (ASGC) in Taiwan, as well as the Advanced Research Computing Services (ARCS) at the University of Virginia. 
\end{acknowledgments}
	
	\nocite{*}
	\bibliographystyle{myapsrev4-2}
	\bibliography{RMFT_vtx-bch1.bib}

\begin{thebibliography}{107}
\bibitem [{not()}]{notezbcp}%
  \BibitemOpen
\href@noop {} {\bibinfo  {journal} {Note that the peak is not exactly at zero
  bias, but still there is a strongly enhanced conductance peak at very low
  energy. For simplicity, we shall call it ZBCP in this paper}\ }\BibitemShut
  {NoStop}%
\bibitem [{Gij()}]{Gijcomplex}%
  \BibitemOpen
\href@noop {} {\bibinfo  {journal} {{H}ere we notice $u_{i}u_{j}^*$ and
  $v_{i}^*v_{j}$ is generally complex. However, the imaginary part, which has
  some pattern to the magnetic flux, is smaller than the real part by about two
  orders of magnitude. Therefore, the calculation of the spectra only
  considered the real part.}\ }\BibitemShut {NoStop}%
\bibitem [{gap()}]{gapargue}%
  \BibitemOpen
\href@noop {} {\bibinfo  {journal} {We should be careful about the gap at low
  dopant. This does not really represent a gap if the system size goes to
  infinity as verified in 3 Tesla or $96\times96$ calculation}\ }\BibitemShut
  {NoStop}%
\end{thebibliography}
	
\end{document}